\documentclass[12pt]{article}

\usepackage{amsmath,amssymb}
\usepackage[bookmarks=true]{hyperref}
\usepackage[nosort]{cite}

\usepackage{epsfig}
\usepackage{graphicx}

\makeatletter
\let\old@startsection=\@startsection
\renewcommand{\@startsection}[6]{\old@startsection{#1}{#2}{#3}{#4}{#5}{#6\mathversion{bold}}}
\makeatother

\def\eeq{\end{eqnarray}}

\def\p{\partial}

\def\=:{=\hspace{-.7em}\raisebox{1.1ex}{.}\hspace{.1em}\raisebox{-0.2ex}{.} }

\newcommand{\beqn}{\begin{eqnarray}}
\newcommand{\eeqn}{\end{eqnarray}}

\newcommand {\beq}{\begin{eqnarray}}
\newcommand {\eeqq}{\end{eqnarray}}

\newcommand {\Tr}{{\rm Tr}\,}

\newcommand{\tN}{\tilde{N}}
\newcommand{\pt}{\partial}

\def\ntwo{${\mathcal N}=2\;$}

\def\none{${\mathcal N}=1\;$}
\def\ntwot{${\mathcal N}=(2,2)\;$}

\makeatletter \@addtoreset{equation}{section} \makeatother

\makeatletter
\let\old@makecaption=\@makecaption
\def\@makecaption{\small\old@makecaption}
\makeatother

\makeatletter
\def\mr@ignsp#1 {\ifx\:#1\@empty\else #1\expandafter\mr@ignsp\fi}%
\newcommand{\multiref}[1]{\begingroup
\xdef\mr@no@sparg{\expandafter\mr@ignsp#1 \: }%
\def\mr@comma{}%
\@for\mr@refs:=\mr@no@sparg\do{\mr@comma\def\mr@comma{,}\ref{\mr@refs}}%
\endgroup}
\makeatother

\ifx\href\asklfhas\newcommand{\href}[2]{#2}\fi

\begin{document}

\begin{flushright}\footnotesize
\texttt{FTPI-MINN-11/08} \\
\texttt{UMN-TH-2941/11\,\,\,\, April 2}
\vspace{0.5cm}
\end{flushright}

\renewcommand{\thefootnote}{\arabic{footnote}}
\setcounter{footnote}{0}
\begin{center}%
{\Large\textbf{\mathversion{bold}
Effective World-Sheet Theory for Non-Abelian Semilocal Strings in \ntwo
Supersymmetric QCD
}
\par}

\vspace{0.8cm}%

\textsc{M. Shifman$^{1}$, W. Vinci$^{1}$ and A. Yung$^{1,2}$  }

\vspace{6mm}
$^1$\textit{William I. Fine Theoretical Physics Institute, University of Minnesota, \\%
 Minneapolis, MN 55455, USA}
\\
\vspace{.3cm}
$^2$\textit{Petersburg Nuclear Physics Institute, Gatchina, St. Petersburg 188300, Russia}

\thispagestyle{empty}
%


\textbf{Abstract}\vspace{5mm}

\begin{minipage}{12.7cm}
We consider non-Abelian semilocal strings (vortices, or vortex-strings) 
arising in $\mathcal N=2$ supersymmetric U$(N)$ gauge theory with 
$N_f=N+\tilde N$ matter hypermultiplets in the fundamental representation (quarks),
and a Fayet--Iliopoulos term $\xi$. We present, for the first time ever,
a systematic  {\em field-theoretic} derivation of the world-sheet theory for such strings,
describing dynamics of both, orientational and size zero modes. Our derivation is complete in the limit 
$( \ln L  )\to \infty$ where $L$ is an infrared (IR) regulator in the transverse plane. In this limit the world-sheet theory is
obtained exactly. It is presented by a so far unknown $\mathcal N=2$ two-dimensional 
sigma model, to which we refer as the $zn$ model, with or without twisted masses.
Alternative formulations of the  $zn$ model are worked out: conventional and extended gauged 
formulations and
a geometric formulation.  We compare the exact metric of the  $zn$ model with that
of the weighted CP$(N_f-1)$ model conjectured by Hanany and Tong, through D-branes,  as the world-sheet theory 
for the non-Abelian 
semilocal strings. The Hanany--Tong set-up has no 
parallel for the field-theoretic IR parameter and metrics of the weighted CP$(N_f-1)$ model and $zn$ model
are different. Still  their quasiclassical excitation spectra coincide.

\end{minipage}

\vspace*{\fill}

\end{center}

\newpage

\tableofcontents

\newpage

\section{Introduction}
\label{iintro}

The exact results obtained in the mid 1990s transformed 
a class of $\mathcal N=2$ supersymmetric gauge theories   
into powerful benchmark models allowing one to  study, to an extent, 
non-perturbative physics of real QCD \cite{Seiberg:1994rs,Seiberg:1994aj}. 
In the last decade we witnessed an enormous progress in the study of 
supersymmetric solitonic objects in the same type of theories 
\cite{Shifman:2007ce,Eto:2006pg,Tong:2005un,Konishi:2007dn}. 
While one usually constructs solitons in a weakly coupled (Higgsed) regime, 
it is possible to use supersymmetry to infer the role of solitons at strong coupling. 
For example, Seiberg and Witten proved that confinement in pure $\mathcal N=2$ 
supersymmetric QCD (SQCD) slightly deformed by a
mass term of the adjoint field
is due  a dual Meissner effect (dual superconductivity):
the chromoelectric charges are confined by flux tubes which form due  to the
monopole condensation \cite{Seiberg:1994rs,Seiberg:1994aj,Seiberg:1994pq}. This mechanism
was anticipated by Nambu, 't Hooft and Mandelstam in the mid 1970s 
\cite{Nambu:1974zg,'tHooft:1975pu,Mandelstam:1974pi}.

Certainly, the most interesting discovery in this range of questions 
is the non-Abelian string (also referred to as vortex-string, or just vortex) 
\cite{Hanany:2003hp,Auzzi:2003fs,Shifman:2004dr,Hanany:2004ea}, 
see also \cite{Shifman:2007ce,Eto:2006pg,Tong:2005un,Tong:2008qd} 
for a review. It generalizes the long-known Abrikosov--Nielsen--Olesen (ANO) 
string \cite{Abrikosov:1956sx,Nielsen:1973cs}:  internal moduli 
describing  the orientation of the chromomagnetic flux in the non-Abelian
group appear on the string world sheet.  Thus, the
non-Abelian strings are the bridge that connects solitons appearing in the Coulomb 
phase with those present  in the Higgs phase. Moreover, they provide  
a physical explanation \cite{Shifman:2004dr,Hanany:2004ea} of 
remarkable correspondences between two-dimensional sigma models 
and four-dimensional gauge theories 
observed previously \cite{Tong:2003pz,Dorey:1998yh,Dorey:1999zk}. 

Non-Abelian strings were first discovered in \ntwo  SQCD with 
the gauge group U$(N)$ and $N_f=N$ flavors 
of fundamental matter hypermultiplets (quarks) 
\cite{Hanany:2003hp,Auzzi:2003fs,Shifman:2004dr,Hanany:2004ea}.
Internal dynamics of the orientational zero  modes of the non-Abelian string 
supported by this theory is described by two-dimensional 
\ntwot supersymmetric CP$(N-1)$ model living on the string 
world sheet \cite{Hanany:2003hp,Auzzi:2003fs,Shifman:2004dr,Hanany:2004ea}.
This result was obtained  both by a straightforward field-theoretic derivation  and 
D-brane-based arguments, see \cite{Shifman:2007ce} for a review. 
More recently, similar non-Abelian strings were constructed and studied 
in a more general class of theories \cite{Eto:2008yi}, 
including SO$(N)$ and Usp$(N)$ gauge theories \cite{Eto:2009bg}, 
and models with arbitrary matter content \cite{Eto:2009bz}.

When one considers theories with a large number of flavors 
(i.e. $N_f=N+\tN > N$), the non-Abelian strings one deals with
are essentially of the semilocal type:  in addition to
translational and orientational moduli, 
they acquire some moduli related to their physical size. 
Semilocal strings are interesting because they interpolate between the ANO-type strings
(at vanishing size)  and sigma-model lumps (at large sizes) 
\cite{Vachaspati:1991dz,Hindmarsh:1991jq,Hindmarsh:1992yy,Preskill:1992bf,Achucarro:1999it}. 
Revealing low-energy dynamics of the semilocal strings
we are able to understand their role in non-perturbative physics of the 
bulk theory. For example, arbitrary thickness of the  semilocal strings may be 
responsible for lost  confinement \cite{Shifman:2006kd}. This issue is related to 
the semilocal string stability, a property which  is not ensured by topology. 
One has to carefully check 
this stability explicitly \cite{Hindmarsh:1991jq,Leese:1992hi,Auzzi:2008wm}.

Derivation of the low-energy effective theory on (non-Abelian) semilocal strings was carried out in the
past in the framework of string theory,  through a D-brane set-up \cite{Hanany:2003hp,Hanany:2004ea}. 
The effective theory was identified as a particular type of a linear gauged sigma
model with an appropriate matter content, namely two-dimensional
CP$(N_f-1)$  with $N$ positive and $\tN$ negative charges (the so-called weighted 
CP model). The latter seems to be a 
natural generalization of the  CP$(N-1)$ model 
appearing on the world sheet at $N=N_f$
to the case $N_f>N$ \cite{Hanany:2003hp,Hanany:2004ea}.

 This construction, known as the K\"ahler quotient, is similar to the well-known 
Atiyah--Drinfeld--Hitchin--Manin
(ADHM) construction for instantons \cite{Atiyah:1978ri}. 
Unfortunately, contrary to the instanton case, the K\"ahler quotient construction for 
strings is unable, in principle, to describe the correct metric on the moduli space. 
This is the reason why an honest and direct derivation from field theory 
{\em per se} is not only desirable, but, in fact, necessary.

This program started in 2006 \cite{Shifman:2006kd}, with further advances ensuing shortly,
in \cite{Eto:2006uw,Eto:2007yv}, by virtue of a more general formalism 
known as {\em moduli matrix}. In these two works the problem was addressed in the limit  of 
the large vortex size, in which the differential BPS equations are reducible to an algebraic system.

In this paper, we undertake a new field-theoretic
calculation of the low-energy effective action for a 
single  non-Abelian semilocal string,
describing dynamics of both, orientational and size zero modes. 
Our derivation is complete in the limit 
$( \ln L  )\to \infty$ where $L$ is an infrared (IR) regulator in the transverse plane. 
In physical terms $L$ is implemented through the (s)quark mass difference,
$L=|\Delta m|^{-1}$.
In this limit the world-sheet theory is
obtained exactly. It is presented by a so far unknown $\mathcal N=2$ two-dimensional 
sigma model, to which we refer as the $zn$ model, with or without twisted masses.
The bosonic part of the action of the ${\mathcal N}= (2,2)$ $zn$ model (without twisted masses) is
\beq
S_{\rm eff}=\int d^2 x\left\{ |\pt_k(z_jn_i)|^2 + \frac{4\pi}{g^{2}} 
\Big[|\partial_{k}n_i|^2+(n^{*}_i \pt_k n_i)^{2} \Big]
\right\},
\label{exactzn}
\eeq
where 
$i=1,...,N$, while $j=1,...,\tN$. The complex fields $n_i$
are subject to the constraint
$$
\sum_{i=1}^N n^{*}_i  n_i =1\,.
$$
The latter is familiar from the CP$(N-1)$ models. 
The additional complex fields $z_j$, descendants of the size moduli,  are unconstrained.
As we will see later, this novel model has rich dynamics.

Alternative formulations of the  $zn$ model are worked out: conventional and extended gauged 
formulations as well as a geometric formulation. {\em En route},
we clarify the disagreement between two works  mentioned
above \cite{Shifman:2006kd,Auzzi:2008wm}. We also explicitly calculate, 
for the first time,  corrections to the metric  in 
inverse powers of the vortex size. 

Needless to say, the effective action (\ref{exactzn}) collects only terms quadratic in derivatives.
As such, it is valid for low-energy excitations. At energies on the world sheet  $\sim |\Delta m|$ 
higher-derivative terms will become important. We will always limit ourselves to the
two-derivative terms.

The leading term in the metric contains an infrared divergence
\beq
\ln {L}{\sqrt\xi},
\label{IRlog}
\eeq
regularized by an  IR cutoff $L$, where $\xi$ is the Fayet--Iliopoulos (FI) parameter \cite{Fayet:1974jb}. The logarithmic divergence above is due to
long-range tails of the semilocal string which fall off as {\em powers} of the distance from the string axis 
(in the perpendicular plane) rather than exponentially. The fact that the 
size zero modes of the Abelian semilocal strings  are
logarithmically non-normalizable was noted long ago 
\cite{Narain:1983in,Ward:1985ij,Leese:1992fn}. In the non-Abelian semilocal
strings both the size and orientational moduli become logarithmically 
non-normalizable \cite{Shifman:2006kd}.
One possibility is to replace an infinite-length string by
that of a finite length. This will also regulate the spread of the string
in the transverse plane \cite{Yung:1999du}. As was mentioned
a more convenient and natural IR regularization, which will maintain the
BPS nature of the solution,  can be provided by a  mass difference
$\Delta m\neq 0$ of the (s)quark masses  \cite{Shifman:2006kd}.
In this paper we will keep in mind the latter option, using $L$ as an auxiliary parameter at intermediate stages,
which, eventually, will be traded for 
  $ 1/|\Delta m |$, so that (\ref{IRlog}) becomes
   \beq
\ln \frac{\sqrt\xi}{|\Delta m |}\,.
\label{IRlogp}
\eeq 
We will always assume that 
   \beq
 \frac{\sqrt\xi}{|\Delta m |}\gg 1\,,
\label{IRlogpp}
\eeq 
and, in the second part, the logarithm of the above parameter will be considered to be (arbitrarily) large.

In our derivation we take advantage of the presence of this IR logarithm in the world-sheet
theory. We  extract the most singular terms in the limit 
in which the logarithm
(\ref{IRlogp}) tends to infinity. This allows us to find the exact metric (in the above  limit).
In this way we arrive at the $zn$ model  on the string world sheet.
This model is novel; it was not known so far. We start its investigation and uncover
interesting features. 

Next,  we compare the $zn$ model with the weighted CP$(N_f-1)$ model suggested by
Hanany and Tong \cite{Hanany:2003hp,Hanany:2004ea} as the world-sheet theory. 
First, we explicitly verify that our field-theory result is different from
the string theory prediction: the scalar curvatures for the two metrics (ours and the Hanany--Tong one)
are not the same. Still
   quasiclassical excitation spectra  of two models coincide.

Summarizing, our main task with regards to non-Abelian semilocal strings is two-fold. First, at large $\rho$ 
(where $\rho$ is the size of the string) we derive
the K\"ahler potential on the target space as an expansion in the powers of $1/|\rho|$, keeping the leading and the 
first subleading terms. The limit of large $L$ is not used here. The second task, the central point of our paper, 
 is to use the limit $\ln \frac{\sqrt\xi}{|\Delta m |} \gg 1$
to derive the {\em exact} world-sheet model (which, in this case, corresponds to small $\rho\ll \xi^{-1/2}$).

The organization of the paper is as follows. In Sec. \ref{sec:field} we introduce 
the bulk model, construct the semilocal string and calculate its world-sheet effective action.
We derive the corresponding K\"{a}hler potential, including the first correction in the inverse size of the string.
In Sec. \ref{sec:exact}
we calculate the exact metric of the world-sheet theory in the limit of 
the large IR logarithm. Section \ref{sec:masses} is devoted to nonvanishing
masses introduced  in   and   their impact on the string world-sheet theory.
In Sec. \ref{sec:spectrum} we calculate the quasiclassical spectrum of excitations in
the world-sheet theory. In Sec. \ref{sec:branes} we review the Hanany--Tong world-sheet theory obtained from  D-branes  
and compare it to our field-theory result. The D-brane derivation is blind to infrared logarithms implying a 
model different from the $zn$ model.
In Sec. \ref{sec:duality} we present the world-sheet effective theory below the crossover (i.e. 
at small $\xi$) and compare it with the weighted CP$(N_f-1)$ model. 
Finally, we conclude and summarize our results in Sec. \ref{sec:conc}.

\section{Non-Abelian Semilocal Strings from Field Theory}
\label{sec:field}

In this section we will derive the string world-sheet theory in the limit of large $\rho$, where
$\rho$ is a size modulus, $|\rho |^2\xi \gg 1$. In this limit a natural expansion parameter appears,
namely the one given in Eq.~(\ref{natex}) below. We will use it in calculating the
effective action to the leading and the first subleading order. Later on (in Sec. \ref{sec:exact}) we will relax the 
constraint $|\rho |^2\xi \gg 1$. At first we must introduce our basic bulk model, on which will rely not only in this section, but throughout the paper.

\subsection{The Bulk Theory}

Our starting point is a U$(N)$ gauge theory with extended $\mathcal N=2$ supersymmetry
and $N_f=N+\tilde N$ fundamental hypermultiplets. The bosonic part of the 
model\,\footnote{The complete bosonic sector includes, in addition,  $N_f$ antifundamental multiplets $\tilde q^{A}$. We set them to zero, $\tilde q^{A} = 0$, as they are trivial in the classical configurations  to be discussed below.} is (see e. g.
\cite{Shifman:2007ce})
\begin{eqnarray}
S & = & \int d^{4}x \left\{   \frac1{4 g^{2}}(F^{0}_{\mu\nu})^{2}+\frac1{4 g^{2}}
(F^{a}_{\mu\nu})^{2}+\frac{1}{g^{2}} |\partial_{\mu}\phi^{0}|^{2}+\frac{1}{g^{2}} |D_{\mu}\phi^{a}|^{2}+ |\nabla_{\mu} q^{A}|^{2} +\right. \nonumber \\[2mm]
 & + &  \frac{g^{2}}{2}\left(  \frac{1}{g^{2}} f^{abc}\bar \phi^{b} \phi^{c}+\bar q^{A} T^a q^{A}  \right)^{2 }+\frac{g^{2}}8(|q^{A}|^{2}-N\xi)^{2} +\nonumber \\[2mm]
 & + & \frac12\left| (\phi^{0}\frac{2}{\sqrt{2N}}+\phi^{a}2 T^{a}+\sqrt2 m_{A})q^{A}  \right|^2\,,
 \label{eq:kinkact}
\end{eqnarray}
with:
\begin{equation}
A=1,2, \dots, N_f\quad \quad \nabla_{\mu}=\partial_{\mu}-\frac{i}{\sqrt{2N}}  A^{0}_{\mu}
-i T^a A^{a}_{\mu}\,.
\end{equation}
The real parameter $\xi$ is the Fayet--Iliopoulos (FI) term \cite{Fayet:1974jb}. As we will see shortly, 
a nonvanishing $\xi$ puts the theory into the Higgs phase. Moreover, the superscripts 0 and $a$ refer to the U(1) and SU$(N)$ parts of the gauge group, respectively. For simplicity we choose both gauge couplings to be equal.
This assumption is not necessary and could have been readily lifted, but we prefer to work with a single gauge coupling $g$.
If  the mass parameters $m_{A}$ are taken real, we  can consistently consider the adjoint fields $a^{0}, \, a^{a}$ to be real as well on the  solitonic solutions. The above expression then simplifies,
\begin{eqnarray}
S & = & \int d^{4}x \left\{   \frac1{4 g^{2}}(F^{0}_{\mu\nu})^{2}+\frac1{4 g^{2}}(F^{a}_{\mu\nu})^{2}+\frac{1}{g^{2}} |\partial_{\mu}\phi^{0}|^{2}+\frac{1}{g^{2}} |D_{\mu}\phi^{a}|^{2}+ |\nabla_{\mu} q^{A}|^{2} +\right. \nonumber \\[3mm]
 & + &  \left. \frac{g^{2}}2\left(\bar q^{A} T^{a}q^{A}  \right)^{2 }
+\frac{g^{2}}8(\bar q^{A}q^{A}-N\xi)^{2} +\ \frac12\left| \left(\phi^{0}\frac{2}{\sqrt{2N}}+\phi^{a} 2 T^{a}+\sqrt2 m_{A}\right)q^{A} \right| \right\}.
\nonumber\\
 \label{eq:kinkactred}
\end{eqnarray}
It is convenient to organize all fields into  matrices, of sizes $N\times N$ and $N\times N_f$, respectively,
\begin{equation}
F_{\mu\nu}\equiv F^{0}_{\mu\nu} \frac{{\bf 1}_N}{\sqrt{2N}} +F^{a}_{\mu\nu} T^a, \quad \Phi\equiv\sqrt2\left(\phi^{0}\frac{{\bf 1}_N}{\sqrt{2N}}+\phi^{a} T^a\right), \quad Q\equiv q_{i}^{A}\,.
\end{equation}
Using the notation above, the action (\ref{eq:kinkactred}) can be written in the following compact form:
\beq
S  =  \int d^{4}x \, \Tr\left\{   \frac1{2 g^{2}}F_{\mu\nu}^{2}+\frac{1}{g^{2}} |D_{\mu}\Phi|^{2}+ |\nabla_{\mu} Q|^{2} +\right.   \left.\frac{g^{2}}4(Q \bar Q -\xi)^{2} +\left|\Phi Q+ Q  M  \right|^{2} \right\}\,,
\nonumber\\
\label{eq:kinkreduced}
\eeq
where the square  mass matrix $M$ is defined as
\beq
M_{AB}=\delta_{AB}m_{A}=
\left(
\begin{array}{cccc}
  m_{1} & 0 & \cdots  & 0  \\
  0 & m_{2}  & \cdots & 0  \\
  \vdots & \cdots & \ddots & \vdots \\
  0 & \cdots  & \cdots & m_{N_f}  
\end{array}
\right) .
\eeq
The nonvanishing (s)quark masses  break the $SU(N)_{\rm F}$ flavor symmetry down to $U(1)_{\rm F}^{N-1}$.  Note that we can always absorb a unit mass matrix into a shift of the adjoint field $\Phi$. Then, with no loss of generality, we can always 
set $$\sum_{A=1}^{N}m_{A}=0\,.$$

\vspace{2mm}

This model described in great detail in \cite{Shifman:2007ce}
has a number of isolated  vacua at generic masses. We choose the vacuum where first $N$ quark flavors condense.
 Up to gauge symmetry transformations we have
\beq
\Phi_{0}=-M, \quad Q=\sqrt\xi \,  \left(1_{N},0_{\tN}\right).
\label{vev}
\eeq
This vacuum is 
 invariant under a  ``color-flavor locked'' global  symmetry $H_{\rm C+F},$\footnote{Note that the vacuum is also invariant under an additional $H_{F}={\rm S}({\rm U}(\tilde n_{1})\times \dots\times {\rm U}(\tilde n_{p}))$ flavor symmetry ($\tilde n_{1}+\dots+\tilde n_{p}=\tilde N)$.}
 \beq
H_{\rm C+F}(\Phi)=H_{\rm C}\Phi H^{-1}_{\rm C}= \Phi\,,  \quad H_{\rm C+F}(Q) \equiv H_{\rm C}\,Q\,H_{\rm F}^{-1}, \quad H_{\rm C}= H_{\rm F}\,.
\eeq
The above symmetry plays an important role in the study of the moduli space of solitons. 
It  is determined by the vacuum value of $\Phi$. In the most general case, 
in which some of the mass parameters are degenerate,
it is given by the stabilizer of the adjoint field,
\beq
H_{\rm C+F}={\rm S}({\rm U}(n_{1})\times {\rm U}(n_{2})\dots\times {\rm U}(n_{q})),\quad n_{1}+\dots+n_{q}=N\,.
\label{eq:colflav}
\eeq
Equation (\ref{eq:colflav}) assumes
that there are $q$ {sets} of fields with degenerate masses. The theory has two parameters 
with mass dimension one, $m\sim m_{i}$ and $\sqrt\xi$, while the dynamical scale $\Lambda$ is implicit.\footnote{For convenience we chose the masses $m_{i}$ to be all of the same order $m$.}
For the time being we will impose the constraints
\beq
 m\ll \sqrt\xi\, ,\qquad \Lambda\ll \sqrt\xi\,.
 \label{2cons}
 \eeq
Then the bulk theory is at weak coupling, and we can reliably deal with the (s)quark  masses as  small deformations 
of the world-sheet theory. In this regime, the symmetry breaking pattern reduces to
\beq
{\rm U}(N)_{C}\times {\rm SU}(N)_{F}\stackrel{\sqrt\xi}{\longrightarrow}{\rm SU}(N)_{C+F}\stackrel{m}{\longrightarrow} H_{C+F}.
\label{eq:Higgs}
\eeq
In particular, in the equal-mass limit the global symmetry group of the bulk theory is
\beq
{\rm SU}(N)_{C+F}\times {\rm SU}(\tN)_{F}\times {\rm U}(1)\,,
\label{c+f}
\eeq
broken down to U(1)$^{(N_f-1)}$ by generic quark mass differences.

At the quantum level, the theory develops a strong coupling scale $\Lambda$. The Higgsing at the scale $\sqrt\xi$ freezes 
the one-loop running of the gauge coupling at the value
\beq
\frac{8 \pi^2}{g^{2}}=(N-\tilde N)\ln\frac{\sqrt\xi}{\Lambda}\,.
\label{bulkcoupling}
\eeq
The theory is asymptotically free for $N>\tilde N$, and conformally invariant at $N=\tilde N$. We will assume, in the following $N >\tilde N$. For large values of the FI term, $\xi\gg\Lambda$,  weak coupling regime sets in (complete Higgsing!), 
and we can reliably construct semiclassical vortex solutions.

\subsubsection*{BPS equations}

The first step in the studies of the BPS-saturated strings
 is to consider the set of the first-order differential equations known as 
 the Bogomol'nyi equations \cite{Bogomol'nyi:1975de}, which follow from the
 Bogomol'nyi  completion of the action (\ref{eq:kinkreduced}) \cite{Eto:2006pg,Hanany:2003hp,Auzzi:2003fs,Shifman:2004dr,Hanany:2004ea,Tong:2003pz,Isozumi:2004vg},
\begin{eqnarray}
S & = & \int d^{4}x\, \Tr\bigg\{  \frac1{g^{2}}\left( F_{12} + \frac{g^{2}}2(Q \bar Q -\xi)\right)^{2} +
 \nonumber \\[2mm]
 & + & |\nabla_{1} Q+i \nabla_{2} Q|^{2}  +  \left| \Phi Q+  Q  M  \right|^{2} + \xi\,F_{12} +
 \nonumber \\[2mm]
 & + & \frac{1}{g^{2}}(F_{ik})^{2}+(\nabla_{k} Q)^{*}(\nabla_{k} Q)+\frac{1}{g^{2}}(F_{kl})^{2} \bigg\}\,, \nonumber \\[2mm]
 & & i=1,2,\quad k,l=0,3\,.
 \label{eq:boghiggs}
\end{eqnarray}
In our notation, $i=1,2$ denotes the transverse (with respect to the string) directions, while $k=0,3$ are 
the space-time coordinates on the string world-sheet. 

The Bogomol'nyi equations are obtained, for static solutions, by requiring each positive-definite contributions above to vanish,
\begin{align}
& \nabla_{1} Q+i \nabla_{2} Q  =  0\,  ,\nonumber \\[2mm] 
  & F_{12} + \frac{g^{2}}2(Q \bar Q -\xi)  =  0\,.
 \label{eq:kinkbps}
 \end{align}
 The string tension is given by  the last term in the second line in  (\ref{eq:boghiggs}), the topological term,
 \begin{eqnarray}
 T  = \xi \int d^{2}x\, \Tr  F_{12}= 2 \pi \xi \, n\, ,
 \end{eqnarray}
where $n$ is the quantized magnetic flux, or equivalently,  the total number of strings.

\subsection{Non-Abelian Semilocal Strings: $\tilde N=1$}
\label{22}

In this section we 
will consider the simplest theory which supports semilocal strings, with a single  ``additional" flavor,
$\tilde N =1$. Semilocal strings are present when the set of vacua of the theory is not simply-connected \cite{Preskill:1992bf}. Actually, the correct topological object to examine in connection
with the
semilocal strings is the second homotopy group of the vacuum manifold, 
which, in the present case, is the complex projective space,
\beq
\pi_{2}(\mathcal {M}_{\rm vac})=\pi_{2}\left({\rm CP}(N-1)\right)= \mathbb{Z}\,.
\label{217}
\eeq 
Equation (\ref{217}) is relevant for the extension of the ANO string in the corresponding semilocal string.
The homotopy group in (\ref{217}) 
is the one lying behind the description of lumps in the associated nonlinear sigma-model,
 which arises as the low-energy limit of the theory (\ref{eq:kinkact}). 
 This is the main reason why semilocal strings are similar to lumps 
 \cite{Hindmarsh:1992yy,Eto:2006uw,Eto:2007yv}. As lumps, the 
 semilocal strings have power-law behaviors at large distances, and possess new {\em size} moduli  determining their characteristic thickness. Nevertheless, they still retain their nature of strings (flux tubes),   which is manifest when we send 
 the size moduli to zero. In this limit we recover just the ANO string, with its exponential behavior \cite{Vachaspati:1991dz}.
 
 Topological stability of the non-Abelian strings is due to the fact that 
 \beq
 \mathbb{Z}_N \in {\rm U}(1)\,\,\, {\rm and }\,\,\, \pi_{1}({\rm U}(1)\times {\rm SU}(N)/ \mathbb{Z}_N )= \mathbb{Z}\,.
 \eeq
Much in the same way as the ANO string, they can be elevated to the semilocal strings, 
see  \cite{Shifman:2007ce}. The winding structure inherent to the non-Abelian vortices, in the context of the semilocal strings, is discussed in the subsequent sections, Eq.~(\ref{219}) and below.

\subsubsection{Ansatz}
\label{ansaa}

Our task is to  explicitly  construct a single semilocal string. 
For the time being we will set all   mass parameters to zero. This is the situation when the full color-flavor 
symmetry SU(N)$_{C+F}$ is preserved, and strings develop size moduli.\footnote{We will reintroduce masses in Sec. \ref{sec:masses}} As was shown in \cite{Hanany:2003hp,Auzzi:2003fs,Shifman:2004dr,Hanany:2004ea}, 
we can easily embed the Abelian ANO-type 
string into a larger non-Abelian gauge group to obtain
 the so-called $\mathbb{Z}_N $ string  \cite{Shifman:2007ce}.
This can be done by choosing  the following ansatz for the matter fields \cite{Shifman:2006kd}:
\begin{eqnarray}
Q_{0}& =& 
\left(
\begin{array}{cccc|c}
  \phi_{1}(r) & 0  & 0 &0 & 0 \\
   0 & \ddots   & \vdots & \vdots & \vdots\\  
  \vdots        &  \dots &   \phi_{1}(r) &0 & 0 \\
 0 & \dots & 0  &  \phi_{2}(r)e^{i \theta}  &\phi_{3}(r) 
\end{array}
\right)\,.
\label{219}
\end{eqnarray}
Equation (\ref{219})  corresponds to a special embedding in which
a nontrivial topological winding is provided by the $N$-th flavor. Technically, it is more convenient to work  
in the singular gauge in which  the fields assume the following form:
\begin{eqnarray}
Q_{0}& =& 
\left(
\begin{array}{ccc|c}
  \phi_{1}(r) & 0  & 0 &0 \\
  \vdots & \ddots   & \vdots & \vdots\\  
 0 & \dots & \phi_{2}(r)  &\phi_{3}(r) 
\end{array}
\right)\equiv\nonumber\\[3mm]
&\equiv &
\bigg(\phi_{1}(r)-n_{0}\,n_{0}^{*}(\phi_{1}(r)-\phi_{2}(r)) \,|\, n_{0}\,\phi_{3}(r)\bigg)\,, \nonumber \\
\end{eqnarray}
while the gauge fields are
\begin{eqnarray}
A_{0,i}& =& \epsilon_{ij} \frac{x_j}{r^2} f(r)\left(
\begin{array}{ccc}
 0 & \dots   & 0\\
 \vdots &  \ddots  & \vdots \\
 0 & \dots & 1   
\end{array}
\right) \equiv n_{0}\, n_{0}^{*}\,\epsilon_{ij}\frac{x_{j}}{r^{2}}f(r)\,.
\label{eq:ansatz}
\end{eqnarray}
In the expressions above  we  introduced an $N$-component vector $n_{0}$ transforming in the fundamental representation of the color flavor group $H_{\rm C+F}$,
\begin{equation}
n_{0}= 
\left(
\begin{array}{c}
 0 \\
  \vdots  \\
  0 \\
     1
\end{array}
\right)\,, \quad n  \equiv  H_{\rm C+F} \,n_{0}\,.
\end{equation}

Given the $\mathbb{Z}_N $ string, and
acting on the solution (\ref{eq:ansatz}) with a generic color-flavor transformation, we get the most 
general vortex-string solution in terms of the orientational vector $n$,
\begin{eqnarray}
Q&=&  \bigg(\phi_{1}(r)-n\,n^{*}(\phi_{1}(r)-\phi_{2}(r)) \,|\, n\,\phi_{3}(r)\bigg) \,,  \nonumber \\[3mm]
A_{i} & = & n\, n^{*}\,
\epsilon_{ij}
\frac{x_{j}}{r^{2}}f(r)\,,
\label{eq:semilocsol}
\end{eqnarray}
where the complex $N$-vector $n_i$ is obviously subject to the condition
\beq
|n_i|^2=1.
\label{unitn}
\eeq

\subsubsection{BPS equations and solutions}
\label{bpsequs}

The non-Abelian Bogomol'nyi  equations reduce to those of the Abelian extended Higgs model. 
With the ansatz (\ref{eq:semilocsol}), we get the following set of equations:
\begin{align}
&r \phi'_{1}(r)=0\, ,\nonumber \\[3mm]
&r \phi'_{2}(r)-f(r)\,\phi_{2}(r)=0\,, \nonumber \\[3mm] 
& r \phi_{3}'(r)-(f(r)-1)\phi_{3}(r)=0\,, \nonumber \\[3mm] 
& \frac1r f'(r)+\frac{g^{2}}{2}(\phi_{2}^{2}(r)+|\phi_{3}(r)|^{2}-\xi)=0\,.
\label{eq:BPSabel}
\end{align}
Note that the first and third equations for $\phi_{1}$ and $\phi_{3}$ can be identically solved by
\beq
\phi_{1}(r)=\sqrt\xi, \quad  \phi_{3} = \frac{\rho}{r}\,\phi_{2}\,.
\label{eq:solexact}
\eeq
In the expression above, $\rho$ is a complex modulus  which parametrizes the size of the semilocal string.
The remaining set of two coupled differential equations, then, must be solved numerically, 
since no analytical solution is known to exist. 

Nevertheless, the peculiarity of the system above is that it admits 
regular and smooth solutions in the limit of large gauge coupling
(the so-called sigma model limit),  $g \rightarrow \infty$. It is even more remarkable that the same system can be solved  algebraically  at any finite power in a $1/g^{2}$ expansion. Keeping only the terms of
the  order of $1/g^{2}$, the solution is  


\begin{eqnarray}
\phi_{2}  =  \phi_{2,0}+\frac1{g^2} \delta \phi_{2}& =& \sqrt\xi\frac{r}{\sqrt{r^{2}+|\rho|^{2}}}+\frac1{g^2} \delta \phi_{2}\,, \nonumber \\[3mm]
f      =  f_{0}+\frac1{g^2} \delta f&  =& \frac{|\rho^{2}|}{r^{2}+|\rho|^{2}}+\frac1{g^2} \delta f \,,\nonumber \\[3mm]
    \delta \phi_{2}  =    - \frac{1}{\sqrt\xi}\frac{2 r |\rho|^{2}}{(r^2+|\rho|^{2})^{5/2}} \,,&&  \delta f  =  \frac{8}{\xi}\frac{r^{2}|\rho|^{2}}{(r^2+|\rho|^{2})^{3}}\,.
    \label{eq:1/g^2exact}
\end{eqnarray}
If we analyze more carefully the validity of the power expansion, by imposing the conditions
\beq
\delta\phi_{2}/(g^{2}\phi_{2,0})\ll1\,,\quad \delta f/(g^{2}f_{0})\ll1\,,
\eeq
we find
\beq
\frac{1}{g\sqrt\xi |\rho|}=\frac{\lambda_{\rm loc}}{\lambda_{\rm semi}}\ll1\,,\quad \quad  \lambda_{\rm loc}=\frac{1}{g\sqrt\xi},\quad \lambda_{\rm semi}=|\rho|\,.
\label{eq:expparam}
\eeq
Thus, the correct expansion parameter is   
\beq
1/(g\sqrt\xi |\rho|)\,,
\label{natex}
\eeq
 or the ratio of the semilocal string size  
to the characteristic size of the local string.

\subsubsection{The Effective Action}
\label{effact}

To calculate the effective action  on the string world
sheet, one first must promote the orientational and size moduli to 
fields depending on the world-sheet coordinates $t$ and $z\equiv x_{3}$,
\begin{eqnarray}
n \rightarrow n(t,x_{3}),\quad \rho \rightarrow \rho(t,x_{3})\,.
\end{eqnarray}
In doing so one has to improve the ansatz  (\ref{eq:semilocsol}) by including a nontrivial 
expression for the world-sheet components of the gauge potential \cite{Shifman:2004dr}, namely,
\begin{eqnarray}
A_{k}& = & -i \bigg(   \partial _{k} n \,n^{*}-n \,\partial_{k}n^{*}- 2n\,n^{*}(n^{*} \partial_{k}n)   \bigg) \omega (r)
\nonumber \\[3mm]
&&-i\, n\, n^{*}\bigg(\rho^{*}\partial_{k}\rho-\rho\partial_{k}\rho^{*}+2 |\rho|^{2}(n^{*}
\partial_{k}n)\bigg)\gamma(r) \,,
\label{eq:gaugetimeansatz}\,
\end	{eqnarray}
where we introduced two profile functions 
$\omega(r)$ and $\gamma(r)$, 
to be determined from a minimization procedure. 
Note that expression (\ref{eq:gaugetimeansatz}) is a refined ansatz as compared to the one introduced 
in Refs. \cite{Shifman:2004dr,Shifman:2006kd}, which does not includes the second term 
proportional to $\gamma$.   The resulting expression for the field strength is 
 \begin{eqnarray}
 F_{ik} & = & \partial_{i}A_{k}-\partial_{k}A_{i}-i [A_{i},A_{k}] \nonumber \\[3mm]
     & = & -\partial_{k}(n\,n^{*})\epsilon_{ij}\frac{x_{j}}{r^{2}}f(r)(1-\omega(r)) \nonumber \\[3mm]
     & - &  i \bigg(   \partial _{k} n \,n^{*}-n \,\partial_{k}n^{*}- 2n\,n^{*}(n^{*} \partial_{k}n)   \bigg) \frac{x_{i}}{r}\omega'(r) \nonumber \\[3mm]
     & - & i\, n\, n^{*}\bigg(\rho^{*}\partial_{k}\rho-\rho\partial_{k}\rho^{*}+2 |\rho|^{2}(n^{*}\partial_{k}n)\bigg)\frac{x_{i}}{r}\gamma'(r)	 \nonumber \\[3mm]
     & - & n\,n^{*}\epsilon_{ij}\frac{x_{j}}{r^{2}}\partial_{k}f(r)\,,
\end{eqnarray}
where the prime in $\omega'$ and $\gamma'$ stands for the first derivative with respect to $r$. 

As a second step, we evaluate the action (\ref{eq:boghiggs}) on the semilocal  solution (\ref{eq:semilocsol}), 
in conjunction 
 with (\ref{eq:gaugetimeansatz}).  We will keep the terms quadratic 
 in the time derivatives with respect to  the world-sheet coordinates,
\begin{equation}
\mathcal L_{\rm eff}=\int dx_{1} dx_{2} \,\Tr\left\{  \frac{1}{g^{2}}(F_{ik})^{2}+(\nabla_{k} Q)^{*}(\nabla_{k} Q) \right\}
\,.
\label{233}
\end{equation}

\subsubsection{The Gauge Kinetic Term}
\label{224}

Evaluation of the gauge kinetic term using the above ans\"atze is straightforward, 
\begin{eqnarray}
 \frac{1}{g^{2}}\Tr(F_{ik})^{2}    & = & \frac{1}{g^{2}}\left(\frac{2}{r^{2}}f^{2}(1-\omega)^{2} +2 \omega'^{2}\right) \left[ \partial_{k}n^{*}\partial_{k}n+(\partial_{k}n^{*} n)^{2}\right] \nonumber \\[3mm]
                           & - & \frac{1}{g^{2}}\left(\gamma'^{2} \right)  \left[\rho^{*}\partial_{k}\rho-\rho\partial_{k}\rho^{*}+2 |\rho|^{2}(n^{*}\partial_{k}n)\right]^{2} \nonumber \\[3mm]
                            & + & \frac{1}{g^{2}}\frac{1}{r^{2}}(\partial_{|\rho|^{2}}f)^{2}\left[\partial_{k}|\rho|^{2}\right]^{2}
                            \label{eq:gauge}\,.
\end{eqnarray}

\subsubsection{The Matter Fields}
\label{themf}

Now we pass to the matter fields. Referring the reader to Appendix \ref{sec:useful} for details we present here the
result of a straightforward albeit rather tedious calculation,
\begin{eqnarray}
&& \Tr[(\nabla_{k} Q)^{*}(\nabla_{k} Q)] \nonumber\\[3mm]
 &=& \bigg[2(\sqrt\xi-\phi_{2})^{2}(1-\omega)+\frac{|\rho|^{2}}{r^{2}}|\phi_{2}|^{2}(1-2\omega)+\bigg(\xi+|\phi_{2}^{2}|(1+\frac{|\rho|^{2}}{r^{2}})\bigg)\omega^{2}\bigg] 
\nonumber \\[3mm]
& \times & \big[\partial_{k}n^{*}\partial_{k}n+(\partial_{k}n^{*} n)^{2}\big] 
\nonumber \\[3mm]
&+&  \left[\left(1+\frac{|\rho|^{2}}{r^{2}}\right)( \partial_{|\rho|^{2}}\phi_{2})^{2}+\frac{1}{r^{2}}\phi_{2}\,\partial_{|\rho|^{2}}\phi_{2}\right](\partial_{k}|\rho|^{2})^{2}   
\nonumber \\[3mm]	
&+& \frac{1}{|r|^{2}}|\phi_{2}|^{2}|\partial_{k}\rho+\rho (n^{*}\partial_{k}n)|^{2} + \frac{1}{r^{2}}|\phi_{2}|^{2}(\rho^{*}\partial_{k}\rho-\rho\partial_{k}\rho^{*}+2 |\rho|^{2}\, n^{*}\partial_{k}n)^{2}\gamma \nonumber \\[3mm]
 &-& |\phi_{2}^{2}|(1+\frac{|\rho|^{2}}{r^{2}})(\rho^{*}\partial_{k}\rho-\rho\partial_{k}\rho^{*}+2 |\rho|^{2}\, n^{*}\partial_{k}n)^{2}\gamma^{2}
\bigg\}\,,
\label{eq:matter}
\end{eqnarray}
where we took advantage of the exact (to all orders in $1/g^{2}$) equations (\ref{eq:solexact}).

\subsubsection{Determination of $\omega(r)$ and $\gamma(r)$}
\label{determ}

To determine the profile functions $\omega(r)$ and $\gamma(r)$, we have to minimize the 
expression given by the sum of two pieces (\ref{eq:gauge}) and (\ref{eq:matter}). The 
minimization with respect of $\omega(r)$ was 
performed in Refs. \cite{Shifman:2004dr,Shifman:2006kd}. It gives the following differential equation:
\begin{eqnarray}
  &-&\frac{2}{g^{2}} \omega'' - \frac{2}{g^{2}r} \omega'-\frac{2}{g^{2}r^{2}}f^{2}(1-\omega) + \big(\xi+\phi_{2}^{2}+\frac{|\rho|^{2}}{r^{2}}\phi_{2}^{2}\big) \omega 
   \nonumber \\[2mm]
   &-& (\xi-\phi_{2})^{2}+\frac{|\rho|^{2}}{r^{2}}\phi_{2}^{2}=0\,,
\end{eqnarray}
which is exactly solved by
\begin{eqnarray}
\omega=1-\frac{\phi_{2}}{\sqrt\xi }\,.
\label{eq:omega}
\end{eqnarray}
Minimization with respect to $\gamma$ gives, on the other hand,
\begin{eqnarray}
 & \frac2{g^{2}}\gamma''+\frac2{g^{2}r}\gamma'+\frac{1}{r}\phi_{2}^{2}-2 r \phi_{2}^{2}\big(1+\frac{|\rho|^{2}}{r^{2}}\big)\gamma=0\,.&
 \end{eqnarray}
The equation above is solved algebraically at zeroth order in $1/g^{2}$ by
\begin{equation}
 \nonumber \\
 \gamma=\frac{1}{2}\frac{1}{r^{2}+|\rho|^{2}} +\frac{1}{g^{2}}\delta \gamma\,.
 \label{eq:gamma}
\end{equation}
We do not evaluate explicitly the term  $\delta \gamma$, since it turns out that it  does not contribute, at the same order $1/g^{2}$, to the effective action. 
\\

\subsubsection{$1/(g^{2}\xi |\rho|^{2})$ corrections to the effective action}
\label{1g2}

We now have all ingredients
necessary to calculate the low-energy effective action for the non-Abelian semilocal string,
 up to the order $1/(g^{2}\xi |\rho|^{2})$. By evaluating the action given  by (\ref{eq:gauge}) and (\ref{eq:matter}), 
exploiting the expressions for the fields (\ref{eq:1/g^2exact}) and (\ref{eq:omega}), and  integrating over 
the transverse plane, we arrive at\,\footnote{See Appendix \ref{sec:useful} for  more details.}  
\begin{eqnarray}
\mathcal L_{\rm eff}   & = &  \pi \xi 
\left(\ln \frac {L^{2}}{|\rho|^{2}}\right)
\big|\partial_{k}(\rho \, n)\big|^{2}-\pi\xi|\partial_{k}\rho+\rho\,(n^{*}\partial_{k}n)|^{2}  
\nonumber\\[3mm]
&+ &\frac{2\pi}{g^{2}}\big[\partial_{k}n^{*}\partial_{k}n+(\partial_{k}n^{*} n)^{2}\big]\,.\nonumber \\
\label{eq:effact}
 \end{eqnarray}
The first term explicitly exhibits the infrared divergence mentioned
in Sec.~\ref{iintro}. An IR divergent integral in the perpendicular plane 
is cut off at $L$ at large distances. 
Thus, the large-size constant $L$ is  introduced to keep the integrations over the transverse plane finite. 
This divergent term was first calculated in Ref. \cite{Shifman:2006kd}. In this paper 
we used the very same approach in order
to obtain its correct expression, which is now consistent with the results 
of Refs.  \cite{Eto:2006uw,Eto:2007yv}, obtained through the moduli matrix formalism. 
The last term in Eq.~(\ref{eq:effact}) is a finite contribution to the metric corresponding to the standard 
Fubini--Study metric on CP$(N-1)$. 


\subsubsection{The K\"{a}hler Potential}
\label{kp}

The effective action (\ref{eq:effact})  describes 1/2-BPS 
saturated
solitons preserving  $\mathcal N=(2,2)$ supersymmetry in 1+1 dimensions, on the world sheet. 
As such, the metric of the world-sheet sigma model  must be given by a  K\"{a}hler potential,
\beq
g_{\phi_{l},\phi_{\bar m}}= \partial_{\phi_{l}}\partial_{\phi_{\bar m}} K(\phi_{l},\phi_{\bar m})\,.
\eeq
Now we will establish its form.

To begin with, let us first introduce a set of holomorphic coordinates $b_i$ and $c$ on the target space,
\beq
b_{i}=\frac{n_{i}}{n_{N}}, \quad c=\rho \,n_{N}\,, \qquad i=1, ... , N-1\,,
\label{241}
\eeq
implying that
\beq
 |\rho|^{2}=(1+\sum_{i}|b_{i}|^{2})|c|^{2}\,. 
\eeq
It is not difficult to show, after some algebra, 
that the following K\"{a}hler potential gives the correct metric on the target space:
\begin{eqnarray}
K_{\rm eff}(b_{i},c,\bar b_{i},\bar c)& =& \pi \xi \,\left(1+\sum_{i}|b_{i}|^{2}\right)
|c|^{2} \,\, \ln \frac {L^{2}}{(1+\sum_{i}|b_{i}|^{2})|c|^{2}}\, 
\nonumber \\[3mm]
& + &   \pi \xi (1+\sum_{i}|b_{i}|^{2})|c|^{2}    +    \frac{2\pi}{g^{2}}\ln(1+\sum_{i}|b_{i}|^{2})
\nonumber \\[3mm]
& = &  \pi \xi \,|\rho|^{2} \left(\ln \frac {L^{2}}{|\rho|^{2}}\right) 
+ \pi \xi |\rho|^{2}+ \frac{2\pi}{g^{2}}\ln\left(1+\sum_{i}|b_{i}|^{2}\right)\,.
\nonumber\\
\label{eq:kahlereffect}
\end{eqnarray}
Note that the above expression is invariant under the color-flavor isometry, as it should be,
of course. The first two terms depend only on the physical size $|\rho|$, which is an obvious invariant. The last term, on the other hand, is invariant up to  K\"{a}hler transformations. 

Let us recall here that the K\"{a}hler potential in the case of the {\em local} non-Abelian string, when $\rho=0$ 
takes the form \cite{Hanany:2003hp,Auzzi:2003fs,Shifman:2004dr,Hanany:2004ea,Isozumi:2004vg}
\begin{eqnarray}
K_{\rm eff}(b,0,\bar b, 0) =\frac{4\pi}{g^{2}}\ln\left(1+\sum_{i}|b_{i}|^{2}\right)\,.  
\label{eq:kahlerloc}
\end{eqnarray}
We would like to draw attention to the
difference of a factor two in the coefficients in front of the logarithms in 
the expressions (\ref{eq:kahlereffect}) and (\ref{eq:kahlerloc}). 
These two terms do not have to coincide, 
since they are  
calculated in the opposite limits $\rho \rightarrow\infty$ and $\rho \rightarrow 0$, respectively .

\subsection{$\tilde N>1$}
\label{ss23}

Now we will lift the requirement $\tilde N =1$.
Generalization to a generic number of flavors requires  more algebra, 
but is quite straightforward. One has to introduce $\tilde N$ complex size moduli $ \rho_{j}$, one for for each 
additional flavor.  $\tilde N $  BPS equations for the fields $q^{N+j}$  are now exactly solved by the following ansatz:
\begin{eqnarray}
q^{N+j}=\frac{\rho_{j}}{r} \, n \, \phi_{2}(r), \qquad j=1,\dots, \tilde N\,. 
\end{eqnarray}
The ansatz for  the gauge potential must be also generalized, namely,
\begin{eqnarray}
A_{k}& = & -i \bigg(   \partial _{k} n \,n^{*}-n \,\partial_{k}n^{*}- 2n\,n^{*}(n^{*} \partial_{k}n)   \bigg) \omega (r)
\nonumber \\[3mm]
&&-i\, n\, n^{*}\sum_{j}^{\tilde N}\bigg(\rho_{j}^{*}\partial_{k}\rho_{j}-\rho_{j}\partial_{k}\rho_{j}^{*}+2 \rho_{j} \rho_{j}^{\dagger}(n^{*}\partial_{k}n)\bigg)\gamma(r) 
\label{eq:gaugetimeansatzgen}\,.
\end	{eqnarray}
The total size of the string $|\rho|^{2}$ is now given  by
\begin{equation}
|\rho|^{2}=\sum_{j=1}^{\tilde N}|\rho_{j}|^{2}\,.
\end{equation}
Taking  into account these relatively insignificant modifications, one gets  the effective action in the form
\begin{eqnarray}
\mathcal L_{\rm eff}   & = &   \pi \xi 
\left(\ln \frac{L^{2}}{|\rho|^{2}}\right)
\sum_{j}|\partial_{k}(n \rho_{j})|^{2} -   \pi\xi\frac{1}{|\rho|^{2}}\bigg| \sum_{j}(\rho_{j}^{*}n^{*})
\partial_{k}(n\rho_{j})\bigg|^{2}
\nonumber \\[3mm]
&-&    \frac{2\pi}{g^{2}} \left(\frac{1}{|\rho|^{2}}\sum_{j}|\partial_{k}(n \rho_{j})|^{2} - \frac{1}{|\rho|^{4}}\bigg| \sum_{j}(\rho_{j}^{*}n^{*})\partial_{k}(n\rho_{j})\bigg|^{2} \right) 
 \nonumber \\[3mm]
& + & \frac{4\pi}{g^{2}} \Big[\partial_{k}n^{*}\partial_{k}n+(\partial_{k}n^{*} n)^{2} \Big]\,.
\label{eq:effactgen}
 \end{eqnarray}
\\

\subsubsection*{The K\"{a}hler potential}

Again, it is convenient to introduce a K\"{a}hler potential for (\ref{eq:effactgen}). We introduce a 
 set of holomorphic coordinates $b_i$ and $c_j$, in parallel with (\ref{241}),
\begin{eqnarray}
b_{i}&=&\frac{n_{i}}{n_{N}}, \qquad c_{j}=\rho_{j} \,n_{N} \,,
\nonumber\\[3mm]
 |\rho|^{2}&=& \left(1+\sum_{i}|b_{i}|^{2}\right)\sum_{j}|c_{j}|^{2}\,, 
 \nonumber\\[3mm]
 i&=&1,... , N-1,\qquad j=1,\dots,\tilde N \,.
 \label{bc}
\end{eqnarray}
Then,  in terms of these coordinates we have
\begin{eqnarray}
&&K_{\rm eff}(b_{i},c_{j},\bar b_{i},\bar c_{j}) =   \pi \xi \,|\rho|^{2} \ln \frac {L^{2}}{|\rho|^{2}}+ \pi \xi |\rho|^{2}- \frac{2\pi}{g^{2}}\ln(|\rho|^{2})
\nonumber\\[4mm]
&+& \frac{4\pi}{g^{2}}\ln\left(1+\sum_{i}|b_{i}|^{2}\right)
 \nonumber \\[4mm]
 & = &  \pi \xi \,|\rho|^{2} \ln \frac {L^{2}}{|\rho|^{2}}+ \pi \xi |\rho|^{2}- \frac{2\pi}{g^{2}}\ln\left(\sum_{j}|c_{j}|^{2}\right)+ \frac{2\pi}{g^{2}}\ln\left(1+\sum_{i}|b_{i}|^{2}\right)\,. \nonumber \\
\label{eq:kahlereffectgen}
\end{eqnarray}
\\

\section{Exact World-Sheet Theory}
\label{sec:exact}

In this section we take advantage of the presence of the IR  divergent term $\ln L/|\rho| \gg 1$
in the K\"ahler potential. Considering
the infrared logarithm as a large parameter (in fact, the largest) we relax the condition $\rho g\sqrt{\xi}\gg 1$ and
obtain the {\em exact} metric of the world-sheet theory to the leading order in the IR logarithm.
By saying `exact' we mean that there is no expansion in $1/(|\rho|^2\,\xi)$ in this metric, unlike 
the results in Sec.~\ref{kp} or \ref{ss23}.

Consider first the case $\tN=1$. The world-sheet theory (\ref{eq:effact}) has the form
\beq
S_{\rm eff}=\int d^2 x\left\{ 2\pi \xi \,\ln \frac {L}{|\rho|}\,|\pt_k (\rho\, n_i)|^2 +{\rm finite\; terms}
\right\},
\label{gstructure}
\eeq
where finite terms are those which do not contain the infrared logarithm. We stress that our derivation in  
Sec.~\ref{sec:field} gives us the exact expression in front of the IR logarithm. The reason is that the logarithmically
divergent term in the world-sheet theory comes from the long-range power tails of the string solution
which we know exactly. Corrections to the string solution associated with the string core at $r\sim 1/g\sqrt{\xi}$ (which we do not control) do not produce the infrared divergent terms in the world-sheet action.

Following \cite{Shifman:2006kd} we introduce a  new variable $z$ replacing the $\rho$ modulus
\beq
z=\rho\,\left[2\pi\xi\,\ln {\frac{L}{ |\rho|}}\right]^{1/2}\,.
\label{z}
\eeq
With the {\em logarithmic accuracy}
 we  rewrite the world-sheet theory in terms of this new variable $z$  as
\beq
S_{\rm eff}=\int d^2 x\left\{ |\pt_k(zn_i)|^2 +{\rm finite\; terms}
\right\},
\label{gstructurez}
\eeq
where in the finite terms we have to express $\rho$ in terms of $z$. We will justify  momentarily that
in the path integral the field $z(x)$ is not large. In fact it is of order of one, $z\sim 1$.
Given that the IR logarithm is large this means that $\rho$ is in fact small. 
This means that we can take the limit $\rho\to 0$ in the finite terms in Eqs.~(\ref{gstructure}) and 
(\ref{gstructurez}). With vanishing $\rho$, the semilocal non-Abelian string reduces to the usual
(local) non-Abelian string, for which world sheet theory is given by the 
CP$(N-1)$ model \cite{Hanany:2003hp,Auzzi:2003fs,Shifman:2004dr,Hanany:2004ea}. Thus, we can write
\beq
{\rm finite\; terms}\left|_{\rho\to 0} \to {\rm CP}(N-1)\;{\rm model}\,,
\right.
\label{finiteterms}
\eeq
and the bosonic part of the action of the world-sheet theory takes the form
\beq
S_{\rm eff}=\int d^2 x\left\{ |\pt_k(zn_i)|^2 + \frac{4\pi}{g^{2}} 
\Big[|\partial_{k}n_i |^2+(n^{*}_i \pt_k n_i)^{2} \Big]
\right\} .
\label{exactws}
\eeq
In terms of new variables 
the infrared logarithm disappeared! Now it is clear that typical fluctuations
of the $z$ field are $z\sim 1$. In Sec. \ref{sec:masses} we will introduce 
mass terms in (\ref{exactws}) which will make this observation even more evident.

Equation (\ref{exactws}) presenting a new world-sheet model in the semilocal string problem, 
to replace that of Hanany and Tong, is
one of our main results.

We stress that the only approximation used here is that the infrared logarithm is large,
\beq
\ln{ (L g\sqrt{\xi})} \gg 1.
\label{largeIRlog}
\eeq
In fact, in order to write (\ref{finiteterms}) and (\ref{exactws}) we need  $\rho$ to be much smaller than
the string core, $\rho\ll 1/g\sqrt{\xi}$. In terms of the field $z$ this reduces to
\beq
|z|^2 \ll \frac1{g^2}\,\ln{ (L \,g\sqrt{\xi})},
\label{approx}
\eeq
which is obviously satisfied in the limit $L\to\infty$. 

In Sec. \ref{sec:masses} we will introduce the (s)quark mass
terms and show that in this case the infrared cutoff $L$ is replaced by the inverse of a typical mass difference
$L\to 1/\Delta m$. 
Then, instead of (\ref{approx}) we have
\beq
|z|^2 \ll \frac1{g^2}\,\ln{ \left(\frac{ g\sqrt{\xi}}{\Delta m}\right)}\,.
\label{approxm}
\eeq
The latter condition (\ref{approx}) is still satisfied provided $\Delta m$ is taken small enough. 
The parameter $g\sqrt{\xi}$ determines the size of the string core and should be understood as an
ultraviolet (UV) cutoff for the low-energy effective world-sheet theory
(\ref{exactws}), see for example \cite{Shifman:2007ce}. 

Now, let us generalize (\ref{exactws}) to the case $\tN>1$. Starting with (\ref{eq:effactgen})
and following the same steps which leads us to (\ref{exactws}) we get 
\beq
S_{\rm eff}=\int d^2 x\left\{ |\pt_k(z_jn_i)|^2 + \frac{4\pi}{g^{2}} 
\Big[|\partial_{k}n_i|^2+(n^{*}_i \pt_k n_i)^{2} \Big]
\right\},
\label{exact}
\eeq
where $i=1,...,N$, while $j=1,...,\tN$ and we introduced new fields $z_j$,
\beq
z_j=\rho_j\,\left[2\pi\xi\,\ln {\frac{L}{ |\rho|}}\right]^{1/2}\,.
\label{zj}
\eeq

Eq. (\ref{exact}) is our final result for the effective low energy theory on the world sheet of the non-Abelian semilocal string. Proceeding to $(N+\tN -1)$ complex  independent variables $b_i$,  and $\varphi_j$, 
\beq
b_{i}=\frac{n_{i}}{n_{N}}, \qquad \varphi_{j}=z_{j} \,n_{N}, \qquad i=1,...,(N-1),\qquad j=1,...,\tN
\label{bvarphi}
\eeq
( c.f. (\ref{bc})) we can write down the K\"{a}hler potential for the theory (\ref{exact}) in the form
\begin{eqnarray}
K_{\rm eff}(b_{i},\varphi_{j},\bar{b}_{i},\bar{\varphi}_j) &=& \sum_{j=1}^{\tN}\left(|\varphi_j|^2 +\sum_{i=1}^{N-1}|(\varphi_j b_i)|^2\right)+ \frac{4\pi}{g^{2}}\ln\left(1+\sum_{i=1}^{N-1}|b_{i}|^{2}\right)
 	\nonumber \\[3mm]
&\equiv& |\zeta|^{2}+\frac{4\pi}{g^{2}}\ln\left(1+\sum_{i=1}^{N-1}|b_{i}|^{2}\right)\,,
\label{exactkahler}
\end{eqnarray}
where we  defined  
\beq
|\zeta|^{2}=\sum_{j=1}^{\tN}\left(|\varphi_j|^2 +\sum_{i=1}^{N-1}|(\varphi_j b_i)|^2\right)\,.
\label{313}
\eeq
This K\"{a}hler potential gives us the world-sheet theory written in the geometric formulation in terms of 
$(N+\tN -1)$ unconstrained complex variables. The disadvantage of this geometric formulation is that 
the global SU$(N)$ symmetry
is not manifest much in the same way as for CP${(N-1)}$ model.
For $\tN=1$ the K\"{a}hler potential (\ref{exactkahler}) describes the blow-up of  ${\mathbb C}^N$ at the origin.

In  Sec.~\ref{sec:masses} we will derive the  world sheet theory for the case of unequal quark masses,
rewrite it in terms of a U(1) gauge theory and discuss its perturbative spectrum.


\section{Inclusion of Quark  Masses}
\label{sec:masses}

\subsection{World-Sheet Theory}

Now we assume that the quark mass differences 
$(m_A-m_B)$ are nonvanishing  in the bulk theory (\ref{eq:kinkact}).
This generates a mass-dependent potential on the non-Abelian semilocal string world sheet \cite{Shifman:2006kd}.
In addition, a natural IR cutoff appears which converts (\ref{IRlog}) in (\ref{IRlogp}).

The leading term in this potential contains the IR logarithm,
\beq
V_{\rm eff} = V_{\rm eff}^{{\rm IR-log}} +V_{\rm eff}^{{\rm finite}}\,  .
\label{Vtot}
\eeq
The first term here was calculated in \cite{Shifman:2006kd} in the case $N=2$. We briefly review  this calculation
and then   generalize it to arbitrary $N$.

Consider first $\tN=1$. The IR-logarithmic contribution  to the potential arises from the last
term in the bulk action (\ref{eq:kinkact}) with $A=N+1$,
\beq
\int d^4 x \left|(\Phi +m_{N+1})\,q^{N+1}\right|^2\,,
\label{bulkmassterm}
\eeq
where $\Phi$ can be replaced by its vacuum expectation value (VEV) (\ref{vev}) with the logarithmic accuracy. Substituting
the string solution for the extra flavor $q^{N+1}$ (\ref{eq:semilocsol}) and using (\ref{eq:solexact}) and 
(\ref{eq:1/g^2exact}) we get 
\beq
\int d^4 x \sum_{i=1}^{N}|m_i-m_{N+1}|^2 \,|n_i|^2\, \xi\,\frac{|\rho|^2}{r^2}\,.
\label{logint}
\eeq
The integral over $r$ in the perpendicular plane gives the IR logarithm,
\beq
V_{\rm eff}^{{\rm IR-log}} =2\pi\xi \int d^2 x \ln{\left(\frac{1}{|\Delta m||\rho|}\right)}\,
\sum_{i=1}^{N}|\rho|^2\,|m_i-m_{N+1}|^2 \,|n_i|^2.
\label{VIR}
\eeq
Here we replaced the IR cutoff $L$ with $1/\Delta m$, which is a typical scale of quark mass differences, $\Delta m \sim (m_A-m_B)$. The reason for this  is that at $(m_A-m_B)\neq 0$ the Higgs branch of the theory is lifted and we do not have 
massless squarks in the bulk theory.  All profile functions for the 
string solution are modified at large $r\geq |\Delta m|^{-1}$ acquiring an exponential fall-off 
$\sim \exp{(-\left|\Delta m\right| r)}$ \cite{Shifman:2006kd}.
Using the variable $z$ (\ref{z}) we can rewrite (\ref{VIR}) as
\beq
V_{\rm eff} = \int d^2 x \sum_{i=1}^{N}|z|^2\,|m_i-m_{N+1}|^2 \,|n_i|^2 +V_{\rm eff}^{{\rm finite}}\,\,.
\label{Vtotz}
\eeq

Now we will follow the same logic that lead us to the exact world-sheet
kinetic terms (\ref{exactws}). Namely, to determine the finite part of the potential in (\ref{Vtotz})
 we take the limit
$\rho \to 0$. In this limit the semilocal string reduces to the local non-Abelian string. Its potential
on the world sheet  is given by the twisted mass terms of the 
CP$(N-1)$ model \cite{Shifman:2004dr,Hanany:2004ea}, see also the review \cite{Shifman:2007ce}. 
The result for the logarithmic part can be rewritten in terms of $z$, as was done
in Sec.~\ref{sec:exact}. 
In this way we arrive at
\begin{eqnarray}
V_{\rm eff} &=& \int d^2 x \left\{\sum_{i=1}^{N}|z|^2\,|m_i-m_{N+1}|^2 \,|n_i|^2 
\right.
\nonumber\\
&+&
\left.
\frac{4\pi}{g^2}\,\left[\sum_{i=1}^{N}|m_i-m|^2 \,|n_i|^2 
-\left|\sum_{i=1}^{N}(m_i-m) \,|n_i|^2\right|^2\right]\right\},
\label{V}
\end{eqnarray}
where $m$ is the average of the first $N$ quark masses,
\beq
m\equiv \frac1{N}\,\sum_{i=1}^{N}m_i\,.
\label{m}
\eeq

Generalization of (\ref{V}) to the case $\tN>1$ is straightforward.  Our final result for the 
bosonic action of the world-sheet theory for the non-Abelian semilocal string is
\begin{eqnarray}
S_{\rm eff} &=& \int d^2 x\left\{ |\pt_k(z_jn_i)|^2 + \frac{4\pi}{g^{2}} 
\Big[|\partial_{k}n_i|^2+(n^{*}_i \pt_k n_i)^{2} \Big]
 +|m_i-m_{j}|^2 \,|z_j|^2|n_i|^2 
\right.
\nonumber\\
&+&
\left.
\frac{4\pi}{g^2}\,\left[\sum_{i=1}^{N}|m_i-m|^2 \,|n_i|^2 
-\left|\sum_{i=1}^{N}(m_i-m) \,|n_i|^2\right|^2\right]\right\},
\label{Sexact}
\end{eqnarray}
where $m_j$ ($j=1,...,\tN$) denote masses of the last $\tN$ quarks of the bulk theory.

This theory is exact in the limit of the large IR logarithm in the same sense as in Sec.~\ref{sec:exact}. 
The only approximation we use is the condition
(\ref{approxm}) which is obviously satisfied once $g\sqrt{\xi}$ is considered as an ultraviolet cutoff for the
theory (\ref{Sexact}). In particular, we assume that
\beq
|m_A|\ll g\sqrt{\xi}, \qquad A=1,...,N_f\,.
\label{smallmasses}
\eeq
The model (\ref{Sexact}) has a hidden U(1) gauge (local) symmetry,
\beq
n_i\to e^{i\alpha}\,n_i, \qquad z_j \to e^{-i\alpha}\,z_j
\label{U1gauge}
\eeq
and therefore the number of (real) degrees of freedom is
\beq
2(N+\tN) -1 -1 = 2(N+\tN -1)\,,
\label{numberdf}
\eeq
where we subtracted two degrees of freedom  associated
with the condition (\ref{unitn}), as well as
one U(1) phase (\ref{U1gauge}),
 from the total number of components of $n_i$ and $z_j$.

\subsection{Gauge formulation}
\label{gf}

\ntwot supersymmetric CP$(N-1)$ model can be nicely formulated in terms of a
U(1) gauge theory in the limit of the
strong gauge coupling. In this  limit gauge fields and their superpartners
become auxiliary \cite{Witten:1978bc,Hanany:1997vm}. Following the same line of reasoning
 we 
consider the local symmetry (\ref{U1gauge}) as a gauge symmetry and rewrite the theory (\ref{Sexact})
as
\begin{eqnarray}
S_{\rm eff} &=& \int d^2 x\left\{ \left|\pt_k(z_jn_i)\right|^2 + 
 \left|\nabla_{k} n_{i}\right|^2 
+\frac1{4e^2}F^2_{kl} + \frac1{e^2}\,
\left|\pt_{k}\sigma\right|^2
\right.
\nonumber\\[3mm]
&+&
\left.
|m_i-m_{j}|^2 \,|z_j|^2|n_i|^2 
+\left|\sqrt{2}\sigma+m_i\right|^2 \left|n_{i}\right|^2 
+ \frac{e^2}{2} \left(|n_{i}|^2 -\frac{4\pi}{g^{2}} \right)^2
\right\},
\nonumber\\[4mm]
&& 
i=1,...,N\,,\qquad j=1,...,\tN\,,
\label{Sgauge}
\end{eqnarray}
where the covariant derivatives are defined as 
\beq
 \nabla_{k}=\pt_{k}-iA_{k}\,.
\label{covarder}
\eeq
It is assumed that at the very end we take limit $e^2\to\infty$.

Note, that  we rescale fields $n_i$ and $z_j$ in (\ref{Sgauge}), which leads to the 
following $D$-term condition:
\beq
  |n_i|^2 =\frac{4\pi}{g^{2}}
\label{unitvec}
\eeq
(in the limit $e^2\to\infty$),  instead of (\ref{unitn}). Moreover, in this limit
the gauge field $A_{k}$  and its \ntwo bosonic superpartner $\sigma$ become
auxiliary and can be eliminated,
\beq
A_k=-i\,n_i^{*}\pt_k n_i, \qquad \sqrt{2}\sigma=- \sum_{i}m_i\,|n_i|^2.
\eeq

The global symmetry of the world sheet theory (\ref{Sgauge}) is (the same as in the bulk theory, see (\ref{c+f}))
\beq
{\rm SU}(N)\times {\rm SU}(\tN)\times {\rm U}(1)
\label{globalsym}
\eeq
broken down to U(1)$^{(N_f-1)}$ by the (s)quark mass differences.

\subsection{An Alternative Gauge Formulation}
\label{aagf}

The gauge formulation described in Sec.~\ref{gf} is simple, but it has the disadvantage of including 
a nonstandard kinetic term which is quartic in fields. In this section we propose an alternative formulation 
in terms of a gauged linear sigma model with the
standard kinetic terms which reduce, at low energies, to the models 
(\ref{Sexact}) and (\ref{Sgauge}). The model presented this section can be considered as an UV completion 
of the model (\ref{Sgauge}). This can be achieved at a price of including a potential term.

\subsubsection{$N=2, \, \,\tN=1$}
\label{n2n1}

For the sake of clarity, let us start from the simplest case. We will extend 
the construction to the most general case in Sec.~\ref{n2nn}. The U(1) 
gauged linear sigma model has the following action:
\begin{eqnarray}
S_{\rm eff} &=& \int d^2 x\left\{ \left|\pt_kZ_i\right|^2 + 
 \left|\nabla_{k} n_{i}\right|^2+ \left|\bar\nabla_{k} R\right|^2 
+\frac1{4e^2}F^2_{kl} + \frac1{e^2}\,
\left|\pt_{k}\sigma\right|^2
\right.
\nonumber\\[3mm]
&+&\left|\sqrt{2}\sigma+m_i\right|^2 \left|n_{i}\right|^2 +\left|\sqrt{2}\sigma\right|^2 \left|R\right|^2 
+ \nonumber \\[3mm]
& + & \left.\frac{e^2}{2} \left(|n_{i}|^2-|R|^{2} -\frac{4\pi}{g^{2}} \right)^2+V_{F}(Z_{A},n_{i},R)
\right\},
\nonumber\\[4mm]
&& 
i=1,2.
\label{Sgaugew}
\end{eqnarray}
where the $Z_{i}$ are neutral scalars, the fields
$n_{i}$ have charge $+1$ and the field $R$ have charge $-1$. The gauge covariant derivative acting on $R$ is, thus,
\beq
\bar \nabla_{k}=\p_{k}+i A_{k}\,.
\eeq
 The first term in the third line is the $D$-term required by supersymmetry while the second one
\begin{eqnarray}
V_{F}(Z_{A},n_{i},R_{j})
&=&
|M|^{2}(|Z_{1} n_{2}-Z_{2}n_{1}|^{2}+|n_{i}|^{2}|R|^{2}+|Z_{i}|^{2}|R|^{2})
\nonumber\\[3mm]
&+&
|m_{i}-m|^{2}|Z_{i}|^{2}
\end{eqnarray}
 is a  judiciously chosen $F$-term potential  which comes from the superpotential
\begin{eqnarray}
W_{F}(Z_{i},n_{i},R_{j})=M\left(Z_{1} n_{2}-Z_{2}n_{1}\right)R+\frac12\left(m_{i}-m\right)Z^{2}_{i}\,,
\end{eqnarray}
where $M$ is an auxiliary mass parameter (a UV parameter), to be sent to infinity.
Note that the coefficients $M$ and $m_{i}-m$ act now as complex masses for all  fields, while previously
we introduced the twisted masses $m_{i}$ only for the $n_{i}$ fields. 

We now take the limit
 \beq
 e^2, \,\, M \to\infty\,,
 \eeq
  and integrate  out massive fields (with masses of order $e,\,\,M$). 
  Integrating out the scalar fields we obtain the following vacuum equations:
\begin{eqnarray}
R=0, \quad Z_{1}n_{2}=Z_{2}n_{1}\,.
\label{422}
\end{eqnarray}
Moreover, in the limit above, the $D$-term condition
\beq
  |n_i|^2 -|R|^{2}=|n_i|^2=\frac{4\pi}{g^{2}}
\label{unitvec2}
\eeq
is implemented.\footnote{It is important that $R=0$ in the vacuum, see (\ref{422}).}  The gauge field $A_{k}$  and its $\mathcal N=(2,2)$ bosonic superpartner $\sigma$ become
auxiliary and can be eliminated too,
\begin{eqnarray}
A_k
&=&
-i\,n_i^{*}\pt_k n_i+i\,R^{*}\pt_k R=-i\,n_i^{*}\pt_k n_i\, , 
\nonumber\\[4mm]
 \sqrt{2}\sigma
&=&
 - \frac{\sum_{i}m_{i}|n_i|^2}{\sum_{i}|n_{i}|^{2}+|R|^{2}}=-\sum_{i}m_{i}|n_{i}|^{2}\,.
\end{eqnarray}
Substituting the relations above into (\ref{Sgaugew}) we obtain:
\begin{eqnarray}
S_{\rm eff} &=& \int d^2 x\left\{ \left|\pt_kZ_1\right|^2+\left|\pt_k\left(Z_1\frac{n_{2}}{n_{1}}\right)\right|^2
+\frac{4\pi}{g^{2}} \big[|\partial_{k}n_i|^2+(n^{*}_i \pt_k n_i)^{2} \big] \right. 
\nonumber \\[3mm]
&+&
\frac{4\pi}{g^2}\,\left[\sum_{i=1}^{N}|m_i-m|^2 \,|n_i|^2 
-\left|\sum_{i=1}^{N}(m_i-m) \,|n_i|^2\right|^2\right]
\nonumber\\[3mm]
 &+& \left.  \sum_{i=1}^{N}|m_{i}-m|^{2}|Z_{1}|^{2}+\sum_{i=1}^{N}|m_{i}-m|^{2}\left|Z_1\frac{n_{2}}{n_{1}}\right|^{2}
 \right\},
\label{Sgaugelow}
\end{eqnarray}
which exactly reduces
 to the theory written in (\ref{Sexact}) with the identification
\begin{equation}
Z_{1}\equiv z\, n_{1}
\\ {}\rule{0mm}{5mm}
\end{equation}

\subsubsection{$N$ arbitrary, $\tN>1$}
\label{n2nn}
Essentially the same construction  as in Sec.~\ref{n2n1} can be 
can be carried out in
 the most general case. Consider the following gauged sigma model:
\begin{eqnarray}
S_{\rm eff} &=& \int d^2 x\left\{ \left|\pt_kZ_A\right|^2 + 
 \left|\nabla_{k} n_{i}\right|^2+ \left|\bar\nabla_{k} R_{B}\right|^2 
+\frac1{4e^2}F^2_{kl} + \frac1{e^2}\,
\left|\pt_{k}\sigma\right|^2
\right.
\nonumber\\[3mm]
&+& \left|\sqrt{2}\sigma+m_i\right|^2 \left|n_{i}\right|^2 +\left|\sqrt{2}\sigma\right|^2 \left|R_{B}\right|^2 
+ \nonumber \\
& + & \left.\frac{e^2}{2} \left(|n_{i}|^2-|R_{B}|^{2} -\frac{4\pi}{g^{2}} \right)^2+V_{F}(Z_{A},n_{i},R_{B})
\right\},
\nonumber\\[4mm]
&& 
A=1,...,\tN N\,,\quad i=1,...,N\,,\quad B=1,... ,\tN(N-1)\,.
\label{Sgaugewgen}
\end{eqnarray}
Note that we introduced a large set of new charge-zero fields $Z_{A}$ and negatively charged fields $R_{B}$. 
The theory in Eq.~(\ref{Sgaugewgen})  includes the potential $V_{F}(Z_{A},n_{i},R_{B})$,
\begin{eqnarray}
V_{F}(Z_{A},n_{i},R_{B}) &=&|M|^{2} \sum_{o=1}^{N-1}\sum_{p=1}^{\tN}|Z_{N (p-1)+1}n_{o+1}-Z_{N (p-1)+o+1}n_{1}|^{2}+\nonumber \\
&+&|M|^{2} | \sum_{o=1}^{N-1}\sum_{p=1}^{\tN}Z_{N(p-1)+o+1}\, R_{(N-1)(p-1)+o} |^{2}+\nonumber \\
 &+& |M|^{2}\sum_{o=1}^{N-1}|\sum_{p=1}^{\tN} Z_{N (p-1)+1}\,R_{(N-1)(p-1)+o}   |^{2}+\nonumber \\
 &+& |M|^{2}\sum_{p=1}^{\tN}|\sum_{o=1}^{N-1}n_{o+1}\,R_{(N-1)(p-1)+o}|^{2}+\nonumber \\
 &+& |M|^{2}\sum_{o=1}^{N-1}\sum_{p=1}^{\tN}|n_{1}R_{(N-1)(p-1)+o}|^{2}+\nonumber \\
 &+& \sum_{j=1}^{\tN}\sum_{i=1}^{N}|m_{i}-m_{j}|^{2}|Z_{N (j-1)+i}|^{2}+\nonumber \\
 &+& |M|^{2}\, \sum_B\, |R_B|^2\,.
 \label{eq:genpot}
\end{eqnarray}
This potential is consistent with $\mathcal N=(2,2)$ supersymmetry since it comes  from 
the following superpotential:
\begin{eqnarray}
W_{F}(Z_{A},n_{i},R_{B})&=& M \sum_{o=1}^{N-1}
\sum_{p=1}^{\tN}R_{(N-1)(p-1)+o}\left(Z_{N(p-1)+1} n_{o+1}
\right.
\nonumber \\[3mm]
&-&
\left.
Z_{N(p-1)+o+1}n_{1}\right)+\nonumber \\[3mm]
&+& \frac12\sum_{j=1}^{\tN}\sum_{i=1}^{N}(m_{i}-m_{j})Z^{2}_{N (j-1)+i}
+\frac{1}{2} \,M\, \sum_B\, R_B^2\,.
\nonumber
\\
\end{eqnarray}
After some straightforward but rather tedious
algebra one can show that vanishing of the potential (\ref{eq:genpot}) requires vanishing of all $R$ fields,
\beq
R_{B}=0\,,\quad \forall \,B\,,
\eeq	
and that the nontrivial constraints on the fields are  given by imposing the vanishing of the first line in (\ref{eq:genpot}),
\beq
Z_{N (p-1)+o+1}=Z_{N (p-1)+1}\frac{n_{o+1}}{n_{1}},\quad o=1,\dots,N-1,\quad p=1,\dots,\tN\,.
\eeq
Using the relations above and the  identifications
\beq
Z_{N(j-1)+1}\equiv z_{j}\,n_{1}
\eeq
we arrive at
\begin{eqnarray}
\sum_{A}|\partial_{k} Z_{A}|^{2}& =& \sum_{i,j}|\partial_{k}(z_{j}n_{i})|^{2}, \nonumber \\[3mm]
 \sum_{i,j}|m_{i}-m_{j}|^{2}|Z_{N(j-1)+i}|^{2} &=& \sum_{i,j}|m_{i}-m_{j}|^{2} |z_{j}|^{2}|n_{i}|^{2}\,.\nonumber \\
\end{eqnarray}
This, in conjunction with
the condition $R_{B}=0$ in Eq. (\ref{Sgaugewgen}),  leads us to  (\ref{Sexact}) again.

\section{Quasiclassical spectrum}
\label{sec:spectrum}

In this section we will analyze the vacuum structure and the  mass spectrum  in our
 world-sheet $zn$ theory. 
It is simpler to obtain it from the action (\ref{Sgauge}) written in the gauged formulation.
In this paper we will limit ourselves to the quasiclassical study of the theory
(\ref{Sgauge}) leaving its investigation at the quantum level for future work.
First we will consider perturbative spectrum.

\subsection{Perturbative Spectrum}
\label{pespe}

If all quark masses are different, the theory (\ref{Sgauge}) has $N$ isolated vacua  at 
\beq
\sqrt{2}\sigma=-m_{i_0}, \qquad n_{i_0}=\sqrt{\frac{4\pi}{g^{2}}}, \qquad n_{i\neq i_0}=0, \qquad z_j=0,
\label{vac}
\eeq
where $i_0$  can acquire any value,
 \beq
 i_0=1, ... , N\,, \,\,\, {\rm while} \,\,\, j=1, ... ,\tN\,.
 \label{ij}
 \eeq
 
The above vacua of the world-sheet theory correspond to $N$ elementary non-Abelian strings.
The spectrum of $n_{i\neq i_0}$ and $z_j$ excitations can be read-off from the action (\ref{Sgauge}),
\beq
m_{n_i}= m_i-m_{i_0}, \qquad i\neq i_0, \qquad m_{z_j}= m_j-m_{i_0}.
\label{spectr}
\eeq

Now suppose that one  of the  masses  of the first $N$ quarks 
coincides with another mass of the  last $\tN$ quarks, $m_{j_0}=m_{i_0}$. 
Then the theory develops a noncompact Higgs 
branch growing from the  
vacuum at $\sqrt{2}\sigma=-m_{i_0}$, namely,
\beq
\sqrt{2}\sigma=-m_{i_0}, \quad n_{i_0}=\sqrt{\frac{4\pi}{g^{2}}}, \quad n_{i\neq i_0}=0, \quad z_{j\neq j_0}=0\,,
\qquad z_{j_0}=z_0\,,
\label{Hbranch}
\eeq
where $z_0$ is  an arbitrary complex number. The (real) dimension of this Higgs branch is 
${\rm dim}\,{\mathcal H}=2$.

Although both kinetic and mass terms in (\ref{Sgauge}) acquire a dependence on $z_0$ the masses of 
$n_{i\neq i_0}$ and $z_j$ excitations remain the same, they are  given by (\ref{spectr}). It is only the field $z_{j_0}$
that
becomes massless; it corresponds to fluctuations along the Higgs branch.

Now, let us go to very low energies, much lower than the quark mass differences
$|\Delta m|$. Then, the low-energy effective theory
on the Higgs branch is just a trivial free-field theory for the massless complex field $z_{j_0}$,
\beq
S_{{\rm Higgs\,\, branch}} = \int d^2 x \, |\pt_k z_{j_0}|^2\,.
\label{Hbrtheory}
\eeq

If more than one masses of the
first $N$ quarks coincide with certain masses of
the  last $\tN$ quarks, more noncompact 
Higgs branches
 develop. These Higgs branches are not lifted in quantum theory. 
 In contrast, the compact Higgs branches which classically develop provided that 
 several masses of first quarks coincide with each other
 are lifted in quantum theory much in the same way as in 
CP$(N-1)$ model.

\subsection{Semiclassical kink spectrum}
\label{sec:kinkspectrum}

In addition to perturbative excitations, the theory (\ref{Sgauge}) supports BPS kinks interpolating between different
vacua. Let us calculate their masses in the quasiclassical approximation. To do so we write down the 
Bogomol'nyi
representation for the kink energy. Assuming for simplicity that the
quark masses and $\sigma$ are real and 
that all fields depend only on $x_3$
 we can rewrite (\ref{Sgauge}) in the limit $e^2\to \infty$ as follows:
\begin{eqnarray}
E_{{\rm kink}} &=& \int d x_3\left\{
\rule{0mm}{6mm}
 \left|\pt_{x_3}(z_jn_i)\right|^2 + 
 \left|\nabla_{x_3} n_{i}\right|^2 
+|m_i-m_{j}|^2 \,|z_j|^2|n_i|^2 \right.
\nonumber\\[3mm]
&+&
\left.\left|\sqrt{2}\sigma+m_i\right|^2 \left|n_{i}\right|^2 
\right\}
\nonumber\\[3mm]
&=&
\int d x_3\left\{ \left|\pt_{x_3}(z_jn_i)+(m_i-m_{j})\,z_j\,n_i\right|^2
+\left|\nabla_{x_3} n_{i}+(\sqrt{2}\sigma+m_i)\,n_{i}\right|^2
\right.
\nonumber\\[3mm]
&+&
\left.
 \frac{4\pi}{g^{2}}\,\sqrt{2}\pt_{x_3}\sigma\right\},
\label{bogws}
\end{eqnarray}
where we use the constraint (\ref{unitvec}) and dropped the boundary terms 
\beq
(\sqrt{2}\sigma+m_i)\,|n_{i}|^2\,\,\, {\rm and}\,\,\, 
(m_i-m_{j}) \,|z_j|^2|n_i|^2\,.
\label{56aa}
\eeq
 In particular, the last one vanishes 
 at generic masses in all vacua (\ref{vac}), 
 because $z_j=0$, while on the Higgs branches (\ref{Hbranch}) 
 it is zero because $m_{i_0}=m_{j_0}$.

From the 
Bogomol'nyi representation (\ref{bogws}) 
we see that the kink profile functions satisfy the first-order equations
\begin{eqnarray}
&&
\pt_{x_3}(z_jn_i)+(m_i-m_{j})\,z_j\,n_i=0,
\nonumber\\[3mm]
&&
\nabla_{x_3} n_{i}+(\sqrt{2}\sigma+m_i)\,n_{i}=0,
\label{kfoews}
\end{eqnarray}
while the kink masses  are given by the boundary term in (\ref{bogws}). In particular,
the mass of the kink interpolating between the ``neighboring'' vacua $i_0$ and $i_0+1$ is
\begin{eqnarray}
m^{{\rm kink}}_{i_0\to i_0+1}
&=&
\left|\frac{4\pi}{g^{2}}\,\sqrt{2}\Big[\sigma(x_3=\infty)-\sigma(x_3=-\infty)
\Big]\right|
\nonumber\\[3mm]
&=&
\left|\frac{4\pi}{g^{2}}\,\left(m_{i_0}-m_{i_0+1}\right)\right|  .
\label{kmassws}
\end{eqnarray}

For generic masses the solution of the first-order equations (\ref{kfoews}) is particularly simple. 
The first equation
is solved by $z_j=0$, while the second one reduces to the  first-order equation for BPS kinks 
in the CP$(N-1)$ model with 
twisted masses \cite{Dorey:1998yh}. Thus, the kinks' profile functions are the same as in 
the CP$(N-1)$ model.

We recall that the monopoles are confined in the bulk theory  in the Higgs vacuum (\ref{vev}). In fact, in
the   U$(N)$
gauge theories they are presented by junctions of two  different elementary non-Abelian strings.
Since $N$ elementary non-Abelian strings correspond to $N$ vacua of the world-sheet theory,
the   confined
monopoles of the bulk theory are seen as kinks in the world-sheet 
theory \cite{Tong:2003pz,Shifman:2004dr,Hanany:2004ea}.

As was shown in \cite{Dorey:1998yh},
the BPS spectrum of dyons (at the singular point on the Coulomb branch 
in which  $N$ quarks become massless) in the
four-dimensional  bulk theory  (\ref{eq:kinkact}), for $N_f=N$, identically
coincides with the BPS spectrum in the two-dimensional twisted-mass deformed CP$(N-1)$ model.
 The reason for this coincidence was revealed 
in \cite{Shifman:2004dr,Hanany:2004ea}.
Although the 't Hooft--Polyakov monopole on the Coulomb branch
looks very different from the string junction of the theory in the Higgs regime,
amazingly, their masses are the same 
\cite{Shifman:2004dr,Hanany:2004ea}. This is due to the fact that
the mass of the BPS states (the string junction is a 1/4-BPS state) cannot depend on
$\xi$ because $\xi$ is a nonholomorphic parameter. Since the confined monopole
emerges in the world-sheet theory as a kink, the Seiberg--Witten
formula for its mass should coincide with the exact result for the kink
mass in two-dimensional \ntwo twisted-mass deformed  CP$(N-1)$ model, which is 
the world-sheet theory for the non-Abelian string in the bulk theory with $N_f=N$. 
 Thus, we arrive at
the statement of coincidence of the BPS spectra in both theories.

Clearly the same correspondence should be true also in the $N_f>N$ case. Let us verify
the coincidence of the BPS spectra of the bulk and world-sheet theories in the quasiclassical approximation.
Taking the limit $\xi\to 0$ in (\ref{eq:kinkreduced}) we see that, in the vacuum (\ref{vev}),
the massive gauge bosons and first $N$ quarks have masses
\beq 
m_{N\times N}=m_i-m_{i'}, \qquad i,i'=1,...,N,\qquad i\neq i',
\label{Wcb}
\eeq
while the last $\tN$ quarks  
\beq 
m_{N\times \tN}=m_i-m_{j},\qquad i=1,...,N,\qquad j=1,...,\tN.
\label{eqcb}
\eeq
We see that this spectrum is identical to the perturbative spectrum of the world-sheet theory
(\ref{spectr}).

The monopole spectrum of the bulk theory is given by the Seiberg--Witten formula \cite{Seiberg:1994pq}
\beq
m_{\rm monopole}=|\vec{a}_D\, \vec{n}_m|\approx \left|\frac{4\pi}{g^{2}}\, \vec{a}\, \vec{n}_m\right|,
\label{monspectr}
\eeq
where we use the quasiclassical approximation. Moreover, $\vec{a}$ represents diagonal entries of the adjoint field $\Phi$
while  $\vec{a}_D$  stands for corresponding dual potentials and $\vec{n}_m$ is the magnetic charge of a
 monopole. 
 In particular, for the elementary monopoles $\vec{n}_m= (0,...,1,-1,0,...)$ (with 
nonvanishing entries at the $i$-th and $(i+1)$-th positions)  we get
\beq
m_{\rm monopole}\approx\left|\frac{4\pi}{g^{2}}\,\left(m_{i}-m_{i+1}\right)\right|,\qquad i=1, ... , N-1\,,
\label{monmass}
\eeq
where we use (\ref{vev}).
These masses coincides with the kink masses 
(\ref{kmassws}) of the world-sheet theory in the quasiclassical approximation.
Explicit verification that the exact BPS spectra of both theories agree is left for future work.

\section{Vortices from D-Branes: Comparing with Hanany and Tong}
\label{sec:branes}

\subsection{Weighted CP$(N_f-1)$ model}

As was mentioned in Sec.~\ref{iintro},
non-Abelian semilocal strings were analyzed previously  \cite{Hanany:2003hp,Hanany:2004ea}
within a complementary approach based on $D$-branes.
 To make contact with field theory
it is highly instructive to compare our  field-theoretic results
with those obtained by Hanany and Tong. They 
conjectured that the effective theory on the world sheet of the non-Abelian
semilocal string is given by the weighted CP$(N_f-1)$ model. The latter  can be represented as a 
 strong-coupling limit ($e^2\to\infty$) 
of the  two-dimensional U(1) gauge theory with $N$ positive and $\tN$ negative charges, namely
\begin{eqnarray}
S_{\rm HT} &=& \int d^2 x \left\{
 |\nabla_{k} n_i^{w}|^2 +|\tilde{\nabla}_{k} z_j^w|^2
 +\frac1{4e^2}F^2_{kl} + \frac1{e^2}\,
|\pt_k\sigma|^2
\right.
\nonumber\\[3mm]
&+&
\left|\sqrt{2}\sigma+m_i\right|^2 \left|n_i^{w}\right|^2 
+ \left|\sqrt{2}\sigma+m_{j}\right|^2\left|z_j^w\right|^2
\nonumber\\[3mm]
&+&
\left.
 \frac{e^2}{2} \left(|n_i^{w}|^2-|z_j^w|^2 -\frac{4\pi}{g^2}\right)^2
\right\},
\nonumber\\[3mm]
&& 
i=1,...,N,\qquad j=1,...,\tN\,,\qquad \tilde{\nabla}_k=\pt_k+iA_k\,.
\label{wcp}
\end{eqnarray}
With respect to the U(1) gauge field,
 the fields $n_i^{w}$ and $z_i^w$ have
charges  +1 and $-1$, respectively. We endow these fields with a superscript ``$w$'' (weighted) to distinguish them from 
the $n_i$ and $z_j$ fields which appear in our world-sheet $zn$ theory (\ref{Sgauge}). 
If only the charge $+1$ fields were present, in the limit  $e^2\to\infty$ we would get a conventional twisted-mass deformed
$ {\rm CP}(N-1)$ model.

\subsection{Quasiclassical spectrum}
\label{sec:HTspectrum}

Although the weighted CP$(N_f-1)$
model and the $zn$ model look quite  
different we will show momentarily that the
quasiclassical spectra of excitations of these two models are the same.
Let us start from the perturbative spectrum.

From (\ref{wcp}) we see that the Hanany--Tong world-sheet theory 
has $N$ vacua (i.e. $N$
strings from the standpoint of the bulk theory),
\beq
 \sqrt{2}\sigma=-m_{i_0},\qquad n^w_{i_0}=\sqrt{\frac{4\pi}{g^2}}\,,\qquad n^w_{i\neq i_0}=z^w_j=0\,,
\label{wcpvac}
\eeq
where $i_0=1, ..., N$.

In each vacuum there are $2(N_f-1)$ elementary excitations, counting
real degrees of
freedom, much in the same way as in (\ref{Sgauge}). The action (\ref{wcp}) contains $N$ complex fields
$n^w_i$ and $\tN$ complex fields $z^w_j$.
The  phase of $n^w_{i_0}$ is eaten by the Higgs mechanism.
The condition $|n^w_{i_0}|^2 = \frac{4\pi}{g^2}$ eliminates one extra field.
The physical masses of the elementary excitations
\beq
m_{n^w_i}= m_i-m_{i_0}, \qquad i\neq i_0, \qquad m_{z^w_j}= m_j-m_{i_0}.
\label{HTspectr}
\eeq
This spectrum is identical to the perturbative spectrum of the $zn$ model (\ref{Sgauge})

Now, suppose again that  $m_{j_0}=m_{i_0}$. Then the theory (\ref{wcp}) also develops a noncompact Higgs 
branch growing from the  vacuum at $\sqrt{2}\sigma=-m_{i_0}$, namely
\beq
\sqrt{2}\sigma=-m_{i_0}, \qquad |n^w_{i_0}|^2-|z^w_{j_0}|^2=\frac{4\pi}{g^{2}}, \qquad n^w_{i\neq i_0}=0, 
\qquad z^w_{j\neq j_0}=0.
\label{HTHbranch}
\eeq
 The  (real) dimension of this Higgs branch is two, much in the same way as 
for the
Higgs branch in the world-sheet theory (\ref{Sgauge}). 

Moreover, the spectrum of fields $n^w_{i\neq i_0}$ and $z^w_{j\neq j_0}$ is still given by (\ref{HTspectr}).
One degree of freedom of two complex fields $n^w_{i_0}$ and $z^w_{j_0}$ is eaten by the Higgs mechanism, while the other is fixed by the second constraint in (\ref{HTHbranch}). The remaining two degrees of freedom are massless. They
correspond to fluctuations along the Higgs branch. 

We see that 
the perturbative spectra of these two models (\ref{Sgauge}) and (\ref{wcp}) are identical.

Now consider the effective low-energy theory on the 
Higgs branch (\ref{HTHbranch}). 
At energies below 
the quark mass differences 
only the
fields $n^w_{i_0}$ and $z^w_{j_0}$ are relevant. We resolve the constraint in the second equation in 
(\ref{HTHbranch}) by writing
\beq
n^w_{i_0}=\sqrt{\frac{4\pi}{g^{2}}}\,e^{i\alpha +i\beta}\,\cosh{w},\qquad 
z^w_{j_0}=\sqrt{\frac{4\pi}{g^{2}}}\,e^{i\alpha -i\beta}\,\sinh{w},
\label{constrainthb}
\eeq
where we introduced two phases for two complex fields. From the action (\ref{wcp}) we find the gauge potential
\beq
A_k=2\left(\pt_k\alpha +\frac{\pt_k \beta}{\cosh{2w}}\right).
\eeq
Substituting this together with (\ref{constrainthb}) into the action (\ref{wcp}) we get \cite{Evlampiev:2003ji}
\beq
S^{\rm HT}_{{\rm Higgs \,\, branch}} = \int d^2 x \cosh{2w}\,\Big\{ (\pt_k w)^2+ (\pt_k \beta)^2 \tanh^2 {2w}\Big\},
\label{HTHbtheory}
\eeq
where the common phase $\alpha$ is eaten by the Higgs mechanism, 
and we are left with a sigma model with two real degrees
of freedom.

This theory on the Higgs branch is clearly different from the free theory (\ref{Hbrtheory}). The target space
in (\ref{HTHbtheory}) is  hyperboloid with a nonvanishing curvature. This shows that two models, (\ref{Sgauge}) 
on the one hand and (\ref{wcp}) on the other are different, despite the coincidence of their 
spectra.\footnote{Interrelation between aspects of the Hanany--Tong model and field-theoretic predictions
for non-Abelian strings
was previously studied in \cite{Auzzi:2010jt} in the case of two coaxial strings. There, it was found that a limited number of 
``protected" quantities, such as the BPS spectra, agree, while others (e.g. the metric) disagree.}

Now let us briefly review the kink spectrum of the weighted CP$(N_f-1)$ 
model in the quasiclassical approximation.
Assuming again  the quark masses and $\sigma$ to be real and 
all fields depend only on $x_3$ we cast the 
Bogomol'nyi representation for the kink energy in the model (\ref{wcp}) in the limit $e^2\to \infty$ in the form
\begin{eqnarray}
E_{{\rm kink}} &=& \int d x_3\left\{ 
 \left|\nabla_{x_3} n^w_{i}\right|^2 +\left|\nabla_{x_3} z^w_{j}\right|^2
 +\left|\sqrt{2}\sigma+m_i\right|^2 \left|n^w_{i}\right|^2\right.
 \nonumber\\[3mm]
 &+&
 \left.
\left|\sqrt{2}\sigma+m_j\right|^2 \left|z^w_{j}\right|^2 
\right\}
\nonumber\\[3mm]
&=&
\int d x_3\left\{ \left|\nabla_{x_3} n^w_{i}+(\sqrt{2}\sigma+m_i)\,n^w_{i}\right|^2
+\left|\bar{\nabla}_{x_3} z^w_{j}-(\sqrt{2}\sigma+m_j)\,z^w_{j}\right|^2
\right.
\nonumber\\[3mm]
&+&
\left.
 \frac{4\pi}{g^{2}}\,\sqrt{2}\pt_{x_3}\sigma\right\}.
\label{bogHT}
\end{eqnarray}
This representation shows that kink solutions satisfy the first-order equations
\begin{eqnarray}
&&
\nabla_{x_3} n^w_{i}+(\sqrt{2}\sigma+m_i)\,n^w_{i} =0,
\nonumber\\[3mm]
&&
\bar{\nabla}_{x_3} z^w_{j}-(\sqrt{2}\sigma+m_j)\,z^w_{j}=0,
\label{kfoeHT}
\end{eqnarray}
while the kink masses are given by the boundary term in (\ref{bogHT}). Much in the same way as in 
the theory (\ref{Sgauge}) this gives, for the kink interpolating between vacua $i_0$ and $i_0+1$,
\beq
m^{{\rm kink}}_{i_0\to i_0+1}
=\left|\frac{4\pi}{g^{2}}\,\left(m_{i_0}-m_{i_0+1}\right)\right|.
\label{kmassHT}
\eeq
Again, the kink spectrum 
we get is identical to that in (\ref{kmassws}) obtained in the  world-sheet theory (\ref{Sgauge}).

In Sec.~\ref{vtm} we will show that  geometries of 
the target spaces of these two models are different (in the 
 case when all fields are relevant). Given the agreement of the BPS spectra
this might seem surprising. Maybe not (cf. \cite{Auzzi:2010jt}).  
Such a situation  could have a simple explanation. While the K\"{a}hler potentials of
two \ntwot supersymmetric sigma models are different their
effective twisted superpotentials  
could agree. This would ensure the coincidence of their BPS spectra.

The exact BPS spectrum in 
the weighted CP$(N_f-1)$ model (\ref{wcp}) was originally discussed in \cite{Hanany:1997vm}. 
It was shown to agree
with the BPS spectrum of the bulk theory in the vacuum (\ref{vev}) on the {\em Coulomb branch} (i.e. at $\xi\to 0$)
\cite{Dorey:1999zk}.
This was considered to be a strong argument supporting the 
conjecture that the weighted CP$(N_f-1)$ model (\ref{wcp}) fully
presents a correct world-sheet theory on the semilocal non-Abelian 
string \cite{Hanany:2004ea,Shifman:2006kd}.
Now we are certain that this conjecture is not correct. 
Although the BPS spectrum of weighted CP$(N_f-1)$ model (\ref{wcp})
coincides with that in the bulk theory, this model is different from the genuine world-sheet theory
on the semilocal non-Abelian string, the $zn$ model (\ref{Sgauge}).

\subsection{Comparing two metrics}
\label{vtm}

The K\"{a}hler potential of the theory  (\ref{wcp}) can be written, using the superfield formalism, 
in the following simple form:
\begin{eqnarray}
K_{\rm HT} & = & e^{-V}|n^{w}_{i}|^{2}+e^{V}|z_{j}^{w}|^{2}+\frac{4\pi}{g^{2}}V , 
\nonumber\\[3mm]
i&=&
1,\dots, N, \qquad j=1,\dots,  \tilde N\,,
\end{eqnarray}
where $n^{w}_{i}$ and $z_{j}^{w}$ are chiral superfields and $V$ is a vector superfield and  summations over indices $i$ and $j$ are implicit. We can eliminate  $V$ by solving its equations of motion 
\begin{eqnarray}
\partial_{V }K_{\rm HT} & =&  -e^{-V}|n^{w}_{i}|^{2} +e^{V}|z_{j}^{w}|^{2}+\frac{4\pi}{g^{2}}=0\,;  
 \nonumber \\[3mm]
     & & |n^{w}_{i}|^{2}e^{-2V}-\frac{4\pi}{g^{2}}e^{-V}-|z_{j}^{w}|^{2}=0\,.
\end{eqnarray}
By virtue of the D-term condition, we can assume $|n^{w}_{i}|^{2}\neq0$ whenever $4\pi/g^{2}>0$, then
\begin{eqnarray}
e^{-V} & = & \frac1{2|n^{w}_{i}|^{2}}\left(  \frac{4\pi}{g^{2}}+\sqrt{\left(\frac{4\pi}{g^{2}}\right)^{2}+4 |n^{w}_{i}|^{2}|z_{j}^{w}|^{2}}\,\,\right)\,.
\end{eqnarray}
Substituting this expression back in the K\"{a}hler potential, we obtain, up to K\"{a}hler transformations, the exact expression
\begin{eqnarray}
K_{\rm HT} & = & \frac12\left(  \frac{4\pi}{g^{2}}+\sqrt{\left(\frac{4\pi}{g^{2}}\right)^{2}+4 |M_{ij}|^{2}  }\right)+\frac{2|M_{ij}|^{2} }{  \frac{4\pi}{g^{2}}+\sqrt{\left(\frac{4\pi}{g^{2}}\right)^{2}+4 |M_{ij}|^{2}  }}  \nonumber \\[3mm]
& - & \frac{4\pi}{g^{2}} \ln\left(   \frac{4\pi}{g^{2}}+\sqrt{\left(\frac{4\pi}{g^{2}}\right)^{2}+4 |M_{ij}|^{2}} \right)+\frac{4\pi}{g^{2}} \ln\left(  1+\frac{|M_{i1}|^{2}}{|M_{N1}|^{2}} \right)\,,
\nonumber \\
\label{eq:HTpotential}
\end{eqnarray}
where we defined the meson fields as 
\beq
M_{ij}=n^{w}_{i} z^{w}_{j}.
\eeq
Note that not all of the $N\times\tN$  mesonic fields are independent because of the  relations
 $$M_{ij}M_{kl}=M_{kj}M_{il}\,.$$  
 The total number of independent complex degree of freedoms is $N+\tN-1$, which is 
 the total number of fields in the theory minus one complex rescaling of the fields. We can choose the following set of independent mesons:
\beq
M_{i1}=n^{w}_{i} z^{w}_{1}, \quad M_{Nj}= n^{w}_{N} z^{w}_{j}, \quad i\neq N\,.
\eeq
All other mesons are given by the formula
\beq
M_{ij}=M_{i1}M_{Nj}/M_{N1}\,.
\eeq
The combination $|M_{ij}|^2$ can the be written as
\beq
\sum_{i,j=1}^{N,\tN}|M_{ij}|^2=\sum_{j=1}^{\tN}\left(|M_{Nj}|^{2}+\sum_{i=1}^{N-1}|M_{i1}|^{2}|M_{Nj}|^{2}/|M_{N1}|^{2} \right)\,.
\eeq
To compare the expression above with the 
field-theoretic result (\ref{exactkahler}), we identify the set of independent mesons used in  the K\"{a}hler quotient construction with the set of moduli found in the field-theoretic derivation, 
\beq
 \varphi_{j}\equiv M_{Nj}=n^{w}_{N} z^{w}_{j}, \quad b_{i}\equiv 
\frac{M_{i1}}{M_{N1}}=\frac{n^{w}_{i}}{n^{w}_{N}}, \quad |\zeta|^{2}\equiv|M_{ij}|^{2}\,.
 \eeq
For simplicity, let us compare the two geometries, (\ref{exactkahler}) vs. (\ref{eq:HTpotential}), at  first order 
in the expansion for large $g^{2}$. The K\"{a}hler potential obtained from the Hanany--Tong model is then
\begin{eqnarray}
K_{\rm HT} & = &2 \sqrt{|\zeta|^{2}}-\frac{2\pi}{g^{2}}\log(|\zeta|^{2})+\frac{4\pi}{g^{2}}\log(1+|b_{i}|^{2})\,,
\label{eq:kahlercompht}
\end{eqnarray}
while the exact K\"{a}hler potential we found in field theory is
\begin{eqnarray}
K_{\rm eff}& = & |\zeta|^{2}+\frac{4\pi}{g^{2}}\log(1+|b_{i}|^{2})\,.
\label{eq:kahlercompfield}
\end{eqnarray}
To explicitly demonstrate 
that the two geometries described above are indeed different, 
we calculate the  scalars curvatures
of the respective
target spaces and verify that they disagree. 
For any K\"{a}hler manifold, the Ricci scalar can be easily calculated using the  formulas
\begin{eqnarray}
g_{p\bar q} & = & \p_{p}\p_{\bar q}\, K\,,\nonumber \\[3mm]
R_{p\bar q} & = & -\p_{p}\p_{\bar q} \left(\ln \det g_{p\bar q}\right)\,,\nonumber \\[3mm]
R & = & g^{p \bar q} R_{p\bar q} \,,\qquad p,q=1,N+\tN-1\,,
\label{eq:ricci}
\end{eqnarray}
where  we endow 
the set of independent complex fields which describe the moduli space ($\varphi_{j}$ and $b_{i}$) with indices $p$ and $q$.

Start from the  case $N=2$, $\tN=1$. Evaluating (\ref{eq:ricci}) using  
 (\ref{eq:kahlercompht}) and (\ref{eq:kahlercompfield}) which implies
\beq
|\zeta|^{2}=|\varphi|^{2}(1+|b|^{2})\,,
\eeq
we arrive at
\begin{eqnarray}
R_{\rm HT} & = & \frac{1}{\sqrt{|\zeta|^{2}}}-\frac{2 \pi}{g^{2}|\zeta|^{2}}+\mathcal O(1/g^{2})\,; \nonumber \\[4mm]
R_{\rm eff} & = & 0\,.
\end{eqnarray}
We thus conclude that the geometry of the target space derived from field theory 
has the vanishing Ricci scalar, while for geometry described by the Hanany--Tong model  
the  Ricci scalar does not vanish, rather it falls off as $1/|\zeta|$.

In the case $N=2=\tN=2$, we consider (\ref{eq:kahlercompht}) and (\ref{eq:kahlercompfield}) with 
\beq
|\zeta|^{2}=(|\varphi_{1}|^{2}+|\varphi_{2}|^{2})(1+|b|^{2})\,.
\eeq
The Ricci scalars are then
\begin{eqnarray}
R_{\rm HT} & = & \frac{1}{\sqrt{|\zeta|^{2}}}+\mathcal O(1/g^{2})\,,\nonumber \\[3mm]
R_{\rm eff} & = & -\frac{2}{|\zeta|^{2}}+\frac{8 \pi}{g^{2}|\zeta|^{4}}+\mathcal O(1/g^{2})\,.
\end{eqnarray}
Disagreement is obvious. This parallels the conclusion of \cite{Auzzi:2010jt}.

\section{Duality}
\label{sec:duality}

In this section we will discuss duality relation for the $zn$ model. By no means this relation is
accidental. Rather it is in one-to-one correspondence with the duality relation for the bulk theories.

\subsection{Bulk Duality}
\label{ssbd}

As was shown in \cite{Shifman:2009mb,Shifman:2010id}, at  $\sqrt\xi \sim \Lambda$ 
the bulk theory goes through a crossover transition
to the strong coupling regime. At small  $\xi$ ($\sqrt\xi \ll \Lambda$) this regime can be described
in terms of weakly coupled dual \ntwo SQCD, with the gauge group
\beq
{\rm U}(\tN)\times {\rm U}(1)^{N-\tN}\,,
\label{dualgaugegroup}
\eeq
 and $N_f$ flavors of
light {\em dyons}. 
This non-Abelian \ntwo duality is, in a sense,  similar to Seiberg's duality in \none supersymmetric QCD
\cite{Seiberg:1994pq,Intriligator:1995au}, for further details see
\cite{Shifman:2011ka}. Later a dual non-Abelian gauge group SU$(\tN)$ was identified on the Coulomb branch 
at the root of a baryonic Higgs branch in the \ntwo supersymmetric  SU($N$) gauge theory with massless quarks
\cite{Argyres:1996eh}.

Light dyons are in the fundamental representation of the gauge group
U$(\tN)$ and are charged under Abelian factors in (\ref{dualgaugegroup}). In addition, there are   
light dyons $D^l$ ($l=\tN+1, ..., N$),
 neutral under 
the U$(\tN)$ group, but charged under the
U(1) factors. A small but nonvanishing $\xi$ triggers condensation of all these
dyons,
\beqn
\langle D^{lA}\rangle \! \! &=&\!\!\sqrt{
\xi}\,
\left(
\begin{array}{cccccc}
0 & \ldots & 0 & 1 & \ldots & 0\\
\ldots & \ldots & \ldots  & \ldots & \ldots & \ldots\\
0 & \ldots & 0 & 0 & \ldots & 1\\
\end{array}
\right),
\quad \langle \bar{\tilde{D}}^{lA}\rangle =0,\quad l=1,...,\tN,
\nonumber\\[4mm]
\langle D^{l}\rangle &=& \sqrt{\xi}, \qquad \langle\bar{\tilde{D}}^{l}\rangle =0\,,
\qquad l=\tN +1, ..., N\,.
\label{Dvev}
\eeqn

Now, consider  the equal quark mass case.
Both, the gauge  and flavor SU($N_f$) groups, are
broken in the vacuum. However, the color-flavor locked form of (\ref{Dvev}) guarantees that the diagonal
global SU($\tN)_{C+F}$ survives. More exactly, the  unbroken global group of the dual
theory is 
\beq
 {\rm SU}(N)_F\times  {\rm SU}(\tN)_{C+F}\times {\rm U}(1)\,,
\label{c+fd}
\eeq
the same as in the original theory, see (\ref{c+f}).
Here SU$(\tN)_{C+F}$ is a global unbroken color-flavor rotation, which involves the
last $\tN$ flavors, while SU$(N)_F$ factor stands for the flavor rotation of the 
first $N$ dyons.
Thus, a color-flavor locking takes place in the dual theory too, although in a different way. Now colors are ''locked''
to the last $\tN$ flavors instead of the first $N$, see (\ref{vev}) and (\ref{Dvev}).

 For generic quark masses the  global symmetry  (\ref{c+f}) is broken down to 
U(1)$^{N_f-1}$. 

\subsection{Dual world-sheet  theory}

Much in the same way as 
in the original theory, the presence of the global SU$(\tN)_{C+F}$ group 
is the  reason behind the formation of the non-Abelian strings. We can repeat all the steps that leads us to the 
effective world-sheet theory (\ref{Sgauge}) on the non-Abelian semilocal string 
for the dual bulk theory. Now we have 
$\tN$ orientation moduli $\tilde{n}_j$ with masses $m_j=\{m_{N+1},...,m_{N_f}\}$ 
and $N$ size moduli $\tilde{z}_i$ with masses $m_i=\{m_1,...,m_N \}$ ($j=1,...,\tN$, $i=1, ... , N$).
The bosonic part of the action has the form
\begin{eqnarray}
S^D_{eff} &=& \int d^2 x\left\{ |\pt_k(\tilde{z}_i \tilde{n}_j)|^2 + 
 \left|\nabla_{k} \tilde{n}_j\right|^2 
+\frac1{4\tilde{e}^2}F^2_{kl} + \frac1{\tilde{e}^2}\,
\left|\pt_{k}\sigma\right|^2
\right.
\nonumber\\[3mm]
&+&
\left.
|m_i-m_{j}|^2 \,|\tilde{z}_i|^2|\tilde{n}_j|^2 
+\left|\sqrt{2}\sigma+m_j\right|^2 \left|\tilde{n}_j\right|^2 
+ \frac{\tilde{e}^2}{2} \left(|\tilde{n}_j|^2 -\frac{4\pi}{\tilde{g}^{2}} \right)^2
\right\},
\nonumber\\[4mm]
&& 
i=1,...,N\,,\qquad j=1,...,\tN\,,
\label{Sgauged}
\end{eqnarray}
where $\tilde{g}^{2}$ is the dual bulk coupling, and the strong coupling limit $\tilde{e}\to \infty$
is assumed.

Classically, the vacua of this theory at generic quark masses are at
\beq
\sqrt{2}\sigma=-m_{j_0}, \qquad \tilde{n}_{j_0}=\sqrt{\frac{4\pi}{\tilde{g}^{2}}}, 
\qquad \tilde{n}_{j\neq j_0}=0, \qquad \tilde{z}_i=0,
\label{vacd}
\eeq
where $j_0$  can be $$j_0=1, ... ,\tN\,,$$ while $i=1,...,N$. 
These vacua of the dual world-sheet theory correspond to $\tN$ elementary non-Abelian strings
in the dual bulk theory.

The spectrum of $\tilde{n}_{j\neq j_0}$ and $\tilde{z}_i$ excitations is
\beq
m_{\tilde{n}_{j}}= m_j-m_{j_0}, \qquad j\neq j_0, \qquad m_{\tilde{z}_i}= m_i-m_{j_0}.
\label{spectrd}
\eeq
Note, that this spectrum is different from the perturbative spectrum of the original world-sheet theory,
see (\ref{spectr}).

Suppose again that one  of the  masses 
 of the first $N$ quarks coincides with another mass of the  last $\tN$ quarks, 
 $$m_{j_0}=m_{i_0}\,.$$ Then the dual theory also develops a noncompact Higgs 
branch growing from the  vacuum at $\sqrt{2}\sigma=-m_{j_0}$, namely,
\beq
\sqrt{2}\sigma=-m_{j_0}\,, \quad \tilde{n}_{j_0}=\sqrt{\frac{4\pi}{\tilde{g}^{2}}}\,, \quad \tilde{n}_{j\neq j_0}=0\,,
 \quad \tilde{z}_{i\neq i_0}=0\,,
\quad \tilde{z}_{i_0}=\tilde{z}_0\,,
\label{Hbranchd}
\eeq
where $\tilde{z}_0$ is  a complex number. The (real) dimension of this Higgs branch is 
${\rm dim}{\cal H}=2$.

Again, the masses of the
$\tilde{n}_{j\neq j_0}$ and $\tilde{z}_{i\neq i_0}$ excitations remain the same,  they are 
given in (\ref{spectrd}).  The field $\tilde{z}_{i_0}$
becomes massless, it corresponds to fluctuations along the Higgs branch.

The quasiclassical kink spectrum for the dual world-sheet theory (\ref{Sgauged}) can be
obtained much in the same way as was done for the original world-sheet theory in Sec. \ref{sec:kinkspectrum}.
Writing down a 
Bogomol'nyi representation for the dual model (\ref{Sgauged}) analogous to that in 
(\ref{bogws}) we get the masses of the kinks
 interpolating between the ``neighboring'' vacua $j_0$ and $j_0+1$, see (\ref{vacd}),
\beq
m^{{\rm kink}}_{j_0\to j_0+1}=\left|\frac{4\pi}{\tilde{g}^{2}}\,\sqrt{2}\left[\sigma(x_3=\infty)-\sigma(x_3=-\infty)
\right]\right|
=\left|\frac{4\pi}{\tilde{g}^{2}}\,\left(m_{j_0}-m_{j_0+1}\right)\right|  .
\nonumber\\
\label{kmasswsd}
\eeq

It is straightforward to check that this kink spectrum coincides with the monopole spectrum of 
the dual bulk theory in the quasiclassical approximation.

\subsection{Dual weighted CP($N_f-1$)  model}

Let us start from Hanany and Tong.
The brane-based arguments of \cite{Hanany:2003hp,Hanany:2004ea} can be applied to 
the dual bulk theory too. This leads us to a dual
weighted CP($N_f-1$). Now it has $\tN$ orientational moduli $\tilde{n}^w_j$ with the 
U(1) charge $+1$. 
In addition,  it has $N$ size moduli $\tilde{z}^w_i$ with the U(1) charge $-1$.
The bosonic action of this  model is
\begin{eqnarray}
S^{\rm D}_{\rm HT} \!\!\! &=& \!\!\!\int d^2 x \left\{
 |\nabla_{k} \tilde{n}_j^{w}|^2 +|\tilde{\nabla}_{k} \tilde{z}_i^w|^2
 +\frac1{4\tilde{e}^2}F^2_{kl} + \frac1{\tilde{e}^2}\,
|\pt_k\sigma|^2
\right.
\nonumber\\[3mm]
\!\!&+&\!\!\!\!\left.
\left|\sqrt{2}\sigma+m_j\right|^2 \left|\tilde{n}_j^{w}\right|^2 
+ \left|\sqrt{2}\sigma+m_{i}\right|^2\left|\tilde{z}_i^w\right|^2
+ \frac{\tilde{e}^2}{2} \left(|\tilde{n}_j^w|^2 -|\tilde{z}_i^{w}|^2 -\frac{4\pi}{\tilde{g}^2}\right)^2
\right\},
\nonumber\\[3mm]
&& 
i=1,...,N,\qquad j=1,...,\tN\,.
\label{wcpd}
\end{eqnarray}
It is easy to see that
the classical vacua of this model are at 
\beq
\sqrt{2}\sigma=-m_{j_0}, \qquad \tilde{n}^w_{j_0}=\sqrt{\frac{4\pi}{\tilde{g}^{2}}}, \qquad \tilde{n}^w_{j\neq j_0}=0, \qquad \tilde{z}^w_i=0.
\label{HTvacd}
\eeq

The quasiclassical  spectrum of the dual weighted CP($N_f-1$) model (\ref{wcpd}) can
be obtained along the same lines as in Sec. \ref{sec:HTspectrum}. It appears to be the same as in the dual
$zn$ theory (\ref{Sgauged}), see (\ref{spectrd}) and (\ref{kmasswsd}).

In passing we should mention the following.
It turns out that the weighted CP($N_f-1$) model is selfdual \cite{Eto:2007yv,Shifman:2009mb,Shifman:2010id}.
At $\xi\gg\Lambda^2$ 
the original theory is at weak coupling, and (\ref{bulkcoupling}) is positive. Analytically continuing to the domain
$\xi\ll\Lambda^2$, we formally make it negative, which signals, of course, that the low-energy description
in terms of the original model is inappropriate.
At $\xi \ll \Lambda^2$  the coupling of the infrared free dual bulk theory is given by 
\beq
\frac{8\pi^2}{\tilde{g}^2}(\xi)=
(N-\tN )\ln{\frac{\Lambda}{\sqrt{\xi}}}=-\frac{8\pi^2}{g^2}(\xi)\,.
\label{dbulkcoupling}
\eeq
It becomes positive and the dual model assumes the role of the legitimate low-energy description
(at weak coupling). A direct inspection of 
the dual theory action (\ref{wcpd}) shows that 
the dual theory can be interpreted as a continuation of the sigma model (\ref{wcp})
to  negative values of the coupling constant $g^2$, where we identify
\beq
\tilde{n}^w_{j}=z^w_j,\qquad \tilde{z}^w_i=n^w_i, \qquad i=1,...,N,\qquad j=1,...,\tN.
\label{identfields}
\eeq

\section{Conclusions}
\label{sec:conc}

Our task was to work out an honest-to-god field-theoretic derivation of the world-sheet theory for non-Abelian
semilocal strings. The goal is achieved. The occurrence of the large IR parameter (\ref{IRlogp}) 
not seen in the D-brane derivation proved to be crucial. In the limit when IR logarithm is large 
the world-sheet theory is
obtained exactly. On the string world sheet we discovered a so far unknown $\mathcal N=2$ two-dimensional 
sigma model,  the $zn$ model, with or without twisted masses.
Alternative formulations of the  $zn$ model are worked out: conventional and extended gauged 
formulations and
a geometric formulation.  We compare the exact metric of the  $zn$ model with that
of the weighted CP$(N_f-1)$ model conjectured by Hanany and Tong, through D-branes. 
 In fact these two models 
 are essentially different. This has been unequivocally demonstrated in certain regimes.
Still quasiclassical excitation spectra of two models coincide.
 An obvious task for the future is the large-$N$ solution of the $zn$ model.

\section*{Acknowledgments}
\addcontentsline{toc}{section}{Acknowledgments}
 W.V. thanks Muneto Nitta for the useful discussions and people at Keio University for the kind hospitality while this work was in progress. The work of MS was supported in part by DOE
grant DE-FG02-94ER408. 
The work of W.V. is supported by the DOE grant DE-FG02-94ER40823.
The work of AY was  supported
by  FTPI, University of Minnesota,
by RFBR Grant No. 09-02-00457a
and by Russian State Grant for
Scientific Schools RSGSS-65751.2010.2.

\newpage	

\appendix

\section{Appendix}
\label{sec:useful}

\subsubsection*{A1. Useful formulae}
\addcontentsline{toc}{subsection}{A1. Useful formulae}

For convenience we list here all the relevant traces which appear in the kinetic term for matter fields (\ref{eq:matter}). 
\begin{eqnarray}
&&
\partial_{k}n^{*}\partial_{k}n+(\partial_{k}n^{*} n)^{2}  \equiv  \big[\mathbb C P^{N-1}\big] 
\,,\nonumber \\[2mm]
&&
\Tr\bigg\{\big[\partial_{k}(n\,n^{*})\big]\cdot\big[\partial_{k}(n\,n^{*})\big]\bigg\} =
 2\big[\mathbb C P^{N-1}\big] 
 \,,\nonumber \\[2mm]
&&
\Tr\bigg\{\big[n\,n^{*}\big]\cdot\big[\partial _{k} n \,n^{*}-n \,\partial_{k}n^{*}-2n\,n^{*}(n^{*} \partial_{k}n)\big]\cdot \big[\partial_{k}(n\,n^{*})\big]\bigg\}=
 -\big[\mathbb C P^{N-1}\big] 
\,,\nonumber \\[2mm]
&&
\Tr\bigg\{\big[\partial _{k} n \,n^{*}-n \,\partial_{k}n^{*}-2n\,n^{*}(n^{*} \partial_{k}n)\big]\cdot \big[\partial_{k}(n\,n^{*})\big]\bigg\}=
 0 
 \,,\nonumber \\[2mm]
&&
\Tr\bigg\{\big[n^{*}\big]\cdot\big[\partial _{k} n \,n^{*}-n \,\partial_{k}n^{*}-2n\,n^{*}(n^{*} \partial_{k}n)\big]\cdot \big[\partial_{k}n\big]\bigg\}= -\big[\mathbb C P^{N-1}\big] 
\,,\nonumber \\[2mm]
&&
\Tr\bigg\{\big[\partial _{k} n \,n^{*}-n \,\partial_{k}n^{*}-2n\,n^{*}(n^{*} \partial_{k}n)\big]^{2}\bigg\}=
 -2\big[\mathbb C P^{N-1}\big]  
 \,,\nonumber \\[2mm]
 &&
\Tr\bigg\{\big[n\,n^{*}\big]\cdot\big[\partial _{k} n \,n^{*}-n \,\partial_{k}n^{*}-2n\,n^{*}(n^{*} \partial_{k}n)\big]^{2}\bigg\}= -\big[\mathbb C P^{N-1}\big] \,.
\nonumber \\
\end{eqnarray}

\subsubsection*{A2. Matter fields contributions}
\addcontentsline{toc}{subsection}{A2. Matter fields contributions}

The matter field contribution is evaluated and decomposed in terms of the dependence on powers of the profile functions $\omega$ and $\gamma$

\begin{eqnarray}
\Tr[(\nabla_{k} Q)^{*}(\nabla_{k} Q)]   & =& \mathcal L_{\omega^{0}\gamma^{0}}+\mathcal L_{\omega^{1}}+\mathcal L_{\omega^{2}}+\mathcal L_{\gamma^{1}}+\mathcal L_{\gamma^{2}}\,,
\end{eqnarray}
where
\begin{eqnarray}
\mathcal L_{\omega^{0}\gamma^{0}}&=&\Tr\bigg\{\big[\partial_{k}(\phi_{1}-n\,n^{*}(\phi_{1}-\phi_{2}))\big]\big[\partial_{k}(\phi_{1}-n\,n^{*}(\phi_{1}-\phi_{2}))\big]\bigg\}  
\nonumber\\[2mm]
&+&\big[\partial_{k}(n^{*}\phi_{3}^{*})\big]\big[\partial_{k}(n\,\phi_{3})\big] 
\nonumber \\[2mm]
&=& 2(\phi_{1}-\phi_{2})^{2}\bigg((\partial_{k}n^{*}\partial_{k}n)+(\partial_{k}n^{*} n)^{2}\bigg)+\partial_{k}(n^{*}\phi_{3}^{*})\partial_{k}(n\,\phi_{3}) 
\nonumber \\[2mm]
&+&|\partial_{k}\phi_{1}|^{2}+|\partial_{k}\phi_{2}|^{2} 
\nonumber \\[2mm]
& = & 2(\phi_{1}-\phi_{2})^{2}\big[\mathbb C P^{N-1}\big]+\partial_{k}(n^{*}\phi_{3}^{*})\partial_{k}(n\,\phi_{3})+|\partial_{k}\phi_{1}|^{2}+|\partial_{k}\phi_{2}|^{2}\,,
\nonumber \\[5mm]
\mathcal L_{\omega^{1}}&=&\Tr\bigg\{\big[\phi_{1}-n\,n^{*}(\phi_{1}-\phi_{2})\big]\big[ \partial _{k} n \,n^{*}-n \,\partial_{k}n^{*}-2n\,n^{*}(n^{*}\cdot \partial_{k}n) \big]
\nonumber\\[2mm]
&\times&\big[\partial_{k}(\phi_{1}-n\,n^{*}(\phi_{1}-\phi_{2}))\big]\bigg\}\,\omega
\nonumber \\[2mm]
&+&  \big[n^{*}\phi_{3}^{*}\big]\big[\partial _{k} n \,n^{*}-n \,\partial_{k}n^{*}-2n\,n^{*}(n^{*} \partial_{k}n)\big]\big[\partial_{k}(n\,\phi_{3})\big]\omega +{\rm c.c.}
\nonumber \\[2mm]
 &=&-2\bigg((\phi_{1}-\phi_{2})^{2}+|\phi_{3}|^{2}\bigg)\omega\big[\mathbb C P^{N-1}\big]\,,
 \nonumber \\[5mm]
\mathcal L_{\omega^{2}}&=&-\Tr\bigg\{\big[\phi_{1}-n\,n^{*}(\phi_{1}-\phi_{2})\big]\big[\partial _{k} n \,n^{*}-n \,\partial_{k}n^{*}-2n\,n^{*}(n^{*} \partial_{k}n)\big]^{2}
\nonumber \\[2mm]
&\times& \big[\phi_{1}-n\,n^{*}(\phi_{1}-\phi_{2})\big]\bigg\}\omega^{2}
\nonumber \\[2mm]
&-&  \big[   n^{*}\,\phi_{3}^{*}    \big]\big[\partial _{k} n \,n^{*}-n \,\partial_{k}n^{*}-2n\,n^{*}(n^{*} \partial_{k}n)\big]^{2}\big[    n\,\phi_{3}   \big]\omega^{2}
 \nonumber \\[2mm]
&=& \bigg( 2\phi_{1}^{2}-2(\phi_{1}-\phi_{2})\phi_{2}+(\phi_{1}-\phi_{2})^{2}+|\phi_{3}|^{2}\bigg)
\omega^{2}\big[\mathbb C P^{N-1}\big]
 \nonumber \\[2mm]
&=&\bigg(\phi_{1}^{2}+\phi_{2}^{2}+|\phi_{3}|^{2}\bigg)\omega^{2}\big[\mathbb C P^{N-1}\big]
\,,\nonumber
\end{eqnarray}
\begin{eqnarray}
\mathcal L_{\gamma^{1}}&=&\Tr\bigg\{\big[\phi_{1}-n\,n^{*}(\phi_{1}-\phi_{2})\big]\big[n\,n^{*}\big]
\nonumber\\[2mm]
 &\times&\big[\partial_{k}(\phi_{1}-n\,n^{*}(\phi_{1}-\phi_{2}))\big]\bigg\}\,(\rho^{*}\partial_{k}\rho-\rho\partial_{k}\rho^{*}+2 |\rho|^{2}\, n^{*}\partial_{k}n)\gamma
 \nonumber \\[2mm]
&+&  \big[n^{*}\phi_{3}^{*}\big]\big[\partial_{k}(n\,\phi_{3})\big](\rho^{*}\partial_{k}\rho-\rho\partial_{k}\rho^{*}+2 |\rho|^{2}\, n^{*}\partial_{k}n)\gamma+ {\rm c.c.}
 \nonumber \\[2mm]
 &=&(\phi_{3}^{*}\partial_{k}\phi_{3}-\phi_{3}\partial_{k}\phi_{3}^{*}+2|\phi_{3}|^{2}n^{*}\partial_{k}n)(\rho^{*}\partial_{k}\rho-\rho\partial_{k}\rho^{*}+2 |\rho|^{2}\, n^{*}\partial_{k}n)\gamma
 \,,
 \nonumber \\[5mm]
\mathcal L_{\gamma^{2}}&=&-\Tr\bigg\{\big[\phi_{1}-n\,n^{*}(\phi_{1}-\phi_{2})\big]\big[n\,n^{*}\big]\big[n\,n^{*}\big]\nonumber \\[2mm]
&\times& \big[\phi_{1}-n\,n^{*}(\phi_{1}-\phi_{2})\big]\bigg\}(\rho^{*}\partial_{k}\rho-\rho\partial_{k}\rho^{*}+2 |\rho|^{2}\, n^{*}\partial_{k}n)^{2}\gamma^{2}
\nonumber \\[2mm]
&-&  \big[   n^{*}\,\phi_{3}^{*}    \big]\big[n\,n^{*}\big]\big[n\,n^{*}\big]\big[    n\,\phi_{3}   \big](\rho^{*}\partial_{k}\rho-\rho\partial_{k}\rho^{*}+2 |\rho|^{2}\, n^{*}\partial_{k}n)^{2}\gamma^{2}
\nonumber \\[3mm]
&=&-(\phi_{2}^{2}+|\phi_{3}|^{2})(\rho^{*}\partial_{k}\rho-\rho\partial_{k}\rho^{*}+2 |\rho|^{2}\, n^{*}\partial_{k}n)^{2}\gamma^{2}\,.
\end{eqnarray}
\\

\subsubsection*{A3. Evaluation on the semilocal solution}
\addcontentsline{toc}{subsection}{A3. Evaluation on the semilocal solution}

In this section we explicitly evaluate the integrations over the transverse plane.  We collect all terms appearing in (\ref{eq:gauge}) and (\ref{eq:matter}) in terms of various combinations of derivatives,
\begin{eqnarray}
&& \mathcal L_{\partial_{k}n^{*}\partial_{k}n+(\partial_{k}n^{*} n)^{2}} = 2\pi \int r dr\bigg\{\frac{1}{g^{2}}\left(\frac{2}{r^{2}}f^{2}(1-\omega)^{2} +2 \omega'^{2}\right) 
\nonumber \\[3mm]
&&+ \bigg[2\frac{\phi_{2}}{\sqrt\xi}(\sqrt\xi-\phi_{2})^{2}+\frac{|\rho|^{2}}{r^{2}}|\phi_{2}|^{2}(-1+2\frac{\phi_{2}}{\sqrt\xi})+\bigg(\xi+|\phi_{2}^{2}|(1+\frac{|\rho|^{2}}{r^{2}})\bigg)\left(1-\frac{\phi_{2}}{\sqrt\xi}\right)^{2}\bigg] \bigg\}
\nonumber \\[3mm]
 && \times  \big[\partial_{k}n^{*}\partial_{k}n+(\partial_{k}n^{*} n)^{2}\big]
  \nonumber \\[3mm]
&&  2\pi \int r dr\bigg\{ \xi\left( 1-\frac{|\phi_{2}|^{2}}{\xi}  \right)^{2}  +\frac{|\phi_{2}|^{4}}{\xi}\frac{|\rho|^{2}}{r^{2}}    \bigg\} \big[\partial_{k}n^{*}\partial_{k}n+(\partial_{k}n^{*} n)^{2}\big]
\nonumber \\[3mm]
&&= \frac{2 \pi}{g^{2}}\big[\partial_{k}n^{*}\partial_{k}n+(\partial_{k}n^{*} n)^{2}\big] +2 \pi \xi \ln \frac L{|\rho|}|\rho|^{2}\big[\partial_{k}n^{*}\partial_{k}n+(\partial_{k}n^{*} n)^{2}\big]\,.
\end{eqnarray}
Note that the $1/g^{2}$ corrections drops out from the second piece. Furthermore,

\begin{eqnarray}
&&\mathcal L_{\left[\rho^{*}\partial_{k}\rho-\rho\partial_{k}\rho^{*}+2 |\rho|^{2}(n^{*}\partial_{k}n)\right]^{2}}
\nonumber \\[3mm]
&&= 2\pi \int r dr\bigg\{-\frac{1}{g^{2}}\left(\gamma'^{2} \right) +  \frac{1}{r^{2}}|\phi_{2}|^{2}\gamma- |\phi_{2}^{2}|(1+\frac{|\rho|^{2}}{r^{2}})\gamma^{2}   \bigg\} 
\nonumber \\[3mm]
&&\times[\rho^{*}\partial_{k}\rho-\rho\partial_{k}\rho^{*}+2 |\rho|^{2}\, n^{*}\partial_{k}n]^{2}
 \nonumber \\[3mm]
    &&= 2\pi \int r dr\bigg\{-\frac{1}{g^{2}}\frac{r^{2}}{(r^{2}+|\rho|^{2})^{4}}  +  \frac{1}{4r^{2}(r^{2}+|\rho|^{2})}|\phi_{2}|^{2} \bigg\}
    \nonumber \\[3mm]
&&\times[\rho^{*}\partial_{k}\rho-\rho\partial_{k}\rho^{*}+2 |\rho|^{2}\, n^{*}\partial_{k}n]^{2}
 \nonumber \\[3mm]
&& = \bigg\{ - \frac{\pi}{6g^{2}}\frac{1}{ |\rho|^{4}}+ \frac{\pi \xi}{4} \frac{1}{|\rho|^{2}}-\frac{\pi }{3 g^{2}}\frac{1}{|\rho|^{4}} \bigg\}
\nonumber \\[4mm]
&&\times [\rho^{*}\partial_{k}\rho-\rho\partial_{k}\rho^{*}+2 |\rho|^{2}\, n^{*}\partial_{k}n]^{2}\,, 
\end{eqnarray}
and
\begin{eqnarray}
&&\mathcal L_{\left[\partial_{k}|\rho|^{2}\right]^{2}} 
\nonumber \\[4mm]
&& = 2\pi \int r dr\bigg\{ \frac{1}{g^{2}}\frac{1}{r^{2}}(\partial_{|\rho|^{2}}f)^{2}+\left[\left(1+\frac{|\rho|^{2}}{r^{2}}\right)( \partial_{|\rho|^{2}}\phi_{2})^{2}+\frac{1}{r^{2}}\phi_{2}\,\partial_{|\rho|^{2}}\phi_{2}\right]\bigg\}\big[\partial_{k}|\rho|^{2}\big]^{2}
\nonumber \\[4mm]
 &&=   \bigg\{    \frac{\pi}{6g^{2}}\frac{1}{ |\rho|^{4}}  -\frac{\pi\xi}{4} \frac{1}{|\rho|^{2}}+\frac{ \pi}{3 g^{2}} \frac{1}{|\rho|^{4}} \bigg\}\big[\partial_{k}|\rho|^{2}\big]^{2}\,,
\end{eqnarray}
and
\begin{eqnarray}
\mathcal L_{|\partial_{k}\rho+\rho (n^{*}\partial_{k}n)|^{2}}    & = &2\pi \int r dr    \frac{1}{r^{2}}|\phi_{2}|^{2}|\partial_{k}\rho+\rho (n^{*}\partial_{k}n)|^{2} \nonumber \\
&=& \bigg\{2 \pi \xi \ln \frac L{|\rho|}-\frac{2 \pi }{ g^{2}} \frac{1}{|\rho|^{2}}\bigg\}|\partial_{k}\rho+\rho (n^{*}\partial_{k}n)|^{2}
\end{eqnarray}
Collecting together all the pieces we obtain the result reported in (\ref{eq:effact}), namely,

\begin{eqnarray}
\mathcal L_{\rm eff}   & = &  \pi \xi \ln \frac {L^{2}}{|\rho|^{2}}\big|\partial_{k}(\rho \, n)\big|^{2}-\pi\xi|\partial_{k}\rho+\rho\,(n^{*}\partial_{k}n)|^{2}  
\nonumber \\[2mm]
&+& \frac{2\pi}{g^{2}}\big[\partial_{k}n^{*}\partial_{k}n+(\partial_{k}n^{*} n)^{2}\big] 	\nonumber \\
 \end{eqnarray}


\addcontentsline{toc}{section}{References}

\bibliography{Bibliographysmart}
\bibliographystyle{nb}

\end{document}